\documentclass{elsart}

\usepackage{epsfig}

\usepackage{amssymb}

\newcommand{\st}{$\sigma_{\mathrm{tot}}$ }
\newcommand{\ro}{$\rho$ }
\newcommand{\re}{\mathop{\mathrm{Re}}}
\newcommand{\im}{\mathop{\mathrm{Im}}}

\begin{document}

\begin{frontmatter}



\title{Critical analysis of derivative dispersion relations
at high energies}

\author[ime]{R.F. \'Avila},
\author[ifi]{M.J. Menon}
\address[ime]{Instituto de Matem\'atica, Estat\'{\i}stica e
Computa\c c\~ao Cient\'{\i}fica\\
Universidade Estadual de Campinas, UNICAMP\\
13083-970 Campinas, SP, Brazil}
\address[ifi]{Instituto de F\'{\i}sica Gleb Wataghin\\
Universidade Estadual de Campinas, UNICAMP\\
13083-970 Campinas, SP, Brazil}


\begin{abstract}
We discuss some formal and fundamental aspects related with the
replacement of integral dispersion relations by derivative forms, and
their practical uses in high energy elastic hadron scattering, in
particular $pp$ and $\bar{p}p$ scattering. Starting
with integral relations with one subtraction and considering parametrizations 
for the total cross sections belonging to the class of entire functions in the 
logarithm
of the energy, a series of results is
deduced and our main conclusions are the following: (1) except for the
subtraction constant, the derivative forms do not depend on any
additional free parameter; (2) the only approximation in going from
integral to derivative relations (at high energies) concerns to assume
as zero the lower limit in the integral form; (3) the previous approximation and
the subtraction
constant affect the fit results at both low and high energies and
therefore, the subtraction constant can not be disregarded; 
(4) from a practical point of view,
for single-pole Pomeron and secondary reggeons parametrizations
 and center-of-mass energies above
5 GeV, the derivative relations with the subtraction constant as a free fit 
parameter are completely equivalent to the integral forms with finite (non-zero) 
lower limit. A detailed review on the conditions of validity and assumptions
related with the replacement of integral by derivative relations is also 
presented and discussed.
\end{abstract}

\begin{keyword}
elastic hadron scattering \sep dispersion relations
\sep high energies

\PACS 13.85.Dz \sep 13.85.Lg \sep 13.85.-t
\end{keyword}
\end{frontmatter}


\centerline{\textit{To appear in Nuclear Physics A}}

\newpage

\centerline{\textbf{Contents}}

1. Introduction

2. Historical summary and main points

3. Derivative dispersion relations

\ \ \ \ 3.1 Formal derivation

\ \ \ \ 3.2 Discussion

4. Tests with Pomeron-Reggeon parametrizations

\ \ \ \ 4.1 Basic comments

\ \ \ \ 4.2 Degenerate meson trajectories

\ \ \ \ \ \ \ \ 4.2.1 Analytical results

\ \ \ \ \ \ \ \ 4.2.2 Global fits

\ \ \ \ 4.3 Non-degenerate meson trajectories

\ \ \ \ 4.4 Discussion

5. Conclusions and final remarks

\section{Introduction}

Elastic hadron-hadron scattering, the simplest soft diffractive process,
constitutes one of the hardest challenges in high-energy physics.
Despite all the success of QCD concerning hard and also semi-hard
processes, the large distances involved in the elastic sector (and soft
processes in general) demand a nonperturbative approach and, presently,
we do not know how to calculate an elastic scattering amplitude in a
purely nonperturbative QCD context. At this stage, phenomenology is very
important, but must be based on the constraints imposed on the
scattering amplitudes by some rigorous theorems, deduced from general
principles of the underlying local quantum field theory, namely, Lorentz
Invariance, Unitarity, Analyticity and Crossing.

Dispersion relations play an important role in several areas of Physics,
both as a practical tool and a formal theoretical result. In special,
for particle-particle and particle-antiparticle interactions, they are
consequences of the principles of Analyticity and Crossing
\cite{drgeneral}. In this context, they correlate real and imaginary
parts of crossing even ($+$) and odd ($-$) amplitudes,
which in turn are expressed in terms of the scattering amplitudes for
a given process and its crossed channel, for example, $ a + b $ and
$a + \bar{b} $:

\begin{equation}
F_{ab} = F_{+} + F_{-} \qquad
F_{a\bar{b}} = F_{+} - F_{-}.
\label{eq:1}
\end{equation}

Among the physical quantities that characterize
high-energy elastic hadron-hadron scattering, the total cross section
(optical theorem) and the $\rho$ parameter (related with the phase of
the amplitude) are just expressed in terms of forward real and
imaginary parts of the amplitude \cite{bc},

\begin{equation}
\sigma_{\mathrm{tot}}(s) = \frac{\im F(s,t=0)}{2 k \sqrt s},
\label{eq:2}
\end{equation}

\begin{equation}
\rho(s) = \frac{\re F(s,t=0)}{\im F(s,t=0)},
\end{equation}
where $s$ and $k$ are the center-of-mass energy squared and the
momentum, respectively, and $t$ is the four-momentum transfer squared.
Therefore, a natural and well founded framework for investigating the
behaviors of $\sigma_{\mathrm{tot}}$ and $\rho$, as function of the
energy, is by means of dispersion relations.
On one hand, from analytical parametrizations for \st and fits
to the corresponding experimental data, the \ro parameter may be
determined (analytically or numerically) and compared to the
experimental data and/or extrapolated to higher energies. On the other
hand, \st and \ro may be analytically connected through the dispersion
relations and, therefore, both may be determined from global fits to the
experimental data (that is, simultaneous fits to \st and \ro data),
improving the statistical ensemble in terms of degree of freedom ($F$).

For particles and antiparticles, the $pp$ and $\bar{p}p$ scattering
correspond to the  highest energy interval with available data, reaching
$\sqrt s \sim$ 63 GeV for $pp$ (CERN ISR) and $\sqrt s \sim$ 2 TeV for
$\bar{p}p$ (Fermilab Tevatron). From experiments that are being
conducted at the Brokhaven RHIC it is expected data on $pp$ scattering
at $\sqrt s$: 50 - 500 GeV, and in the near future the CERN LHC will
provide data on $pp$ scattering at 16 TeV. It is highly expected that,
in short term, these novel experiments will allow crucial tests for
several phenomenological models and, among them, the analytical models,
based on the use of dispersion relation techniques, certainly play a
central role.

The use of dispersion relations in the investigation of scattering
amplitudes may be traced back to the end of fifties, when they were
introduced in the form of {\em integral\/} relations. Despite the
important results that have been obtained since then, one limitation of
the integral forms is their non-local character:
in order to obtain the real part of the amplitude, the imaginary part
must be known for all values of the energy. Moreover, the class of
functions that allows analytical integration is limited. By the end of
the sixties and beginning of the seventies, there appeared the
dispersion relations in a {\em differential\/} form,
and they provided new insights in the dispersion relation techniques.
However, as we shall discuss, there still remain some questions related
with both the formal replacement of integral by the derivative forms and
their practical uses in high energies, as, for example, the effects of the
approximations considered, the correct expression of derivative form
(depending on the class of functions involved) and the role of the subtraction
constant. We understand that these aspects must be made clear
before any practical use of the derivative relations.

In this work we critically discuss the situation concerning the
derivative relations and present some answers for the above mentioned
questions. The manuscript is organized as follows. In Section
\ref{sec:historical} we briefly review some historical facts and results
about the integral and derivative relations, stressing the problems we
are interested in. In Section \ref{sec:derivative} we first demonstrate that,
for the class of functions that are entire in the logarithm of the energy,
except for the subtraction constant, the derivative relations do not
depend on any additional free parameter. We then review the main results
in the literature, 
obtained through different approaches, in connection with the corresponding
assumptions and classes of functions involved.
In Section \ref{sec:tests}, making
use of a pomeron-reggeon parametrization, we analyze the practical
effects of the approximations involved in the replacement of the
integral by the derivative forms.
In particular, we show that, for this class of functions, the derivative form with 
the subtraction constant as a free fit parameter is equivalent to the integral 
form for $\sqrt s$ above 5 GeV.
The conclusions and some critical remarks 
are the contents of Section \ref{sec:conclusions}.

\section{Historical summary and main points} \label{sec:historical}

In this Section we briefly review some results related with
the replacement of {\em integral dispersion relations\/} (IDR) by
{\em derivative dispersion relations\/} (DDR), stressing the points that
we shall discuss in the next sections. We are interested in the
high-energy region, specifically $\sqrt s >$ 5 GeV and, as commented
in our introduction, the practical tests will be performed with the $pp$
and  $\bar{p}p$ experimental data on $\sigma_{\mathrm{tot}}$ and $\rho$.
For these scatterings, the experimental data above $\sqrt s \sim$ 20 GeV
indicate a slow increase of $\sigma_{\mathrm{tot}}$, roughly as
$ \sim \ln^2 s$ and that
$\sigma_{\mathrm{tot}}^{\bar{p}p} - \sigma_{\mathrm{tot}}^{pp} \sim 0$
as $s \rightarrow \infty$.
These conditions allow the use of integral dispersion relations with
only one subtraction, and in the standard form they are usually
expressed in terms of the energy $E$ and momentum $p$ of the incoming
proton in the laboratory system \cite{bc,idr}.
Since we shall be interested in the region $\sqrt s > $ 5 GeV, we can
approximate $s = 2m(E + m) \sim 2mE$ with an error $ < 5 \% $, which
decreases as the energy increases.
In this case the standard IDR, with poles removed, are given by

\begin{equation}
\re F_{+}(s)= K + \frac{2s^{2}}{\pi}P\!\!\!\int_{s_{0}}^{+\infty}
\!\!\!\d s'
\frac{1}{s'(s'^{2}-s^{2})}\im F_{+}(s'),
\label{eq:4}
\end{equation}

\begin{eqnarray}
\re F_{-}(s)=  \frac{2s}{\pi}P\!\!\!\int_{s_{0}}^{+\infty} \!\!\!
\d s'
\frac{1}{(s'^{2}-s^{2})}\im F_{-}(s'),
\label{eq:5}
\end{eqnarray}
where $K$ is the subtraction constant, and for $pp$ and $\bar{p}p$
scattering, $s_0=2m^2\sim 1.8$ GeV$^2$. It should be noted that for the
even amplitude the relations with one and two subtractions are equal and
for the odd case the relations without subtraction and with one
subtraction are equal.

Through Eqs. (\ref{eq:1}) to (\ref{eq:5}) the physical quantities \st
and \ro may be simultaneously investigated.
By means of IDR, important results have been obtained since the
beginning of the seventies, as for example in the works by
Bourrely and Fischer \cite{bourrely73}, Amaldi et al. \cite{amaldi77},
Block and Cahn \cite{bc}, Kluit and Timmermans \cite{kluit},
Augier et al. (UA
4/2 Collaboration) \cite{ua42}, Kang, Valin and White  \cite{kvw},
Bertini et al. \cite{bertini} and many others.
However, as commented in our introduction, the inconveniences with the
IDR concern their non-local character and the limited number of
functions that allows analytical integration, and therefore, error
propagation from the fit parameters.
On the other hand, under some conditions, the above integral forms may
be replaced by quasi-local ones, expressed in a derivative form and
called derivative dispersion (or analyticity) relations.  The first
result in that direction appeared indicated in the works by Gribov and Migdal 
\cite{gm},
and afterwards, the DDR were treated in more detail by Bronzan
\cite{bronzan68}, Jackson \cite{jack73}, Bronzan, Kane and Sukhatme
\cite{bks}.
In particular, the forms deduced by the last authors
are given by

\begin{equation}
\re F_{+}(s)  =
s^{\alpha}\tan\left[\frac{\pi}{2}\left( \alpha -1\! +
\frac{\d}{\d\ln s}\right) \right]
\frac{\im F_{+}(s)}{s^{\alpha}},
\end{equation}

\begin{equation}
\re F_{-}(s)  =  s^{\alpha}
\tan\left[\frac{\pi}{2}\left( \alpha  +
\frac{\d}{\d\ln s}\right) \right]
\frac{\im F_{-}(s)}{s^{\alpha}},
\end{equation}
where $\alpha$ is a real parameter. Soon after,
based on the Sommerfeld-Watson-Regge representation and other
assumptions, Kang and Nicolescu introduced the following expression for
the derivative relations \cite{kn}:

\begin{eqnarray}
\frac{\re F_{+}(s)}{s}&=&
\bigg[ \frac{\pi}{2} \frac{\d}{\d\ln s} +
\frac{1}{3} \left(\frac{\pi}{2}\frac{\d}{\d \ln s}\right)^3
\nonumber \\
&+&
\frac{2}{5} \left(\frac{\pi}{2}\frac{\d}{\d \ln s}\right)^5 +
\cdots \bigg]\frac{\im F_{+}(s)}{s},
\label{eq:8}
\end{eqnarray}

\begin{eqnarray}
\frac{\pi}{2}\frac{\d}{\d \ln s}\frac{\re F_{-}(s)}{s} &=&
- \bigg[ 1 - \frac{1}{3}
\left(\frac{\pi}{2}\frac{\d}{\d \ln s}\right)^2
\nonumber \\
& - & \frac{1}{45}
\left(\frac{\pi}{2}\frac{\d}{\d \ln s}\right)^4  - \cdots
\bigg]
\frac{\im F_{-}(s)}{s},
\label{eq:9}
\end{eqnarray}
which, according to the authors, reduces
to the Bronzan, Kane and Sukhatme
result for a decreasing power form of  $\im [F_{-}(s)/s]$ \cite{kn}.
In the last years, these equations have been extensively used
in the detailed analysis on analytical models by Cudell et al. 
\cite{cudelletall} and by the
COMPETE Collaboration \cite{competework}.

The DDR, in the form introduced by {\em Bronzan, Kane and Sukhatme\/} (BKS),
have been criticized by
several authors \cite{crit1,crit2,crit3,crit4}
and analyzed and discussed in detail by Fischer, Kol\'a\v{r} \cite{kf}
and Vrko\v{c} \cite{vrkoc}.
Despite all these works, we understand that two aspects related with the
replacement of the IDR, Eqs. (4) and (5), by the DDR, Eqs. (6), (7),
or (8), (9) and with their practical uses are yet unclear 
in the literature.

One aspect concerns the origin, role and range of the exponent $\alpha$
in the BKS relations, which does not appear in the formulas by Kang and 
Nicolescu.
For example, in the paper by BKS this parameter is allowed to have any real value
\cite{bks}, but in References \cite{crit2,mmp}
(see also \cite{kf}) it must lie in the interval $0 < \alpha < 2$. 
According to some authors, ``the choice of $\alpha$ different from 1
has no practical advantage" \cite{kf} and other authors have used it as 
a free fit parameter \cite{bks,mmpsn}, which demands some physical
interpretation. 
The calculation by Eichmann and Dronkers, in 1974, lead to DDR that do not
depend on $\alpha$ \cite{crit1}, but this parameter appears in a recent 
review by Kol\'a\v{r} and Fischer \cite{kfrev}.
To our knowledge, with the exception of the above quoted considerations
of $\alpha$ as a free fit parameter, all the other practical uses of the
BKS relations are made by assuming $\alpha =$ 1. However, $\alpha$ = 1 
leads to a singularity in the BKS expansion of the odd amplitude, since 
it contains the term $\sec^2(\pi\alpha/2)$ \cite{bks}.
As we shall discuss in detail in Section 3.2, some of these results 
and statements are not contradictories, because they refer either
to different energy regimes (finite or asymptotic $s$) or to
different classes of functions, 
for example, entire functions in the logarithm of the energy, or functions
satisfying general principles from axiomatic field theory, etc... Moreover,
some results have been obtained in different historical contexts in terms
of the highest energies reached in experiments, leading to different
mathematical assumptions on the asymptotic conditions. 

Another aspect concerns the subtraction constant in the singly 
subtracted IDR, Eq. (\ref{eq:4}), which is the starting formula in the
work by Bronzan, Kane and Sukhatme. 
That constant does not appear in all the above derivative forms and, to
our knowledge, neither in almost all their practical uses,
including the detailed works by the COMPETE collaboration.
Yet, to our knowledge, the only exception concerns
the recent works \cite{alm03,lm03,am02}, where we
have performed analysis of the experimental data
available, including cosmic-ray estimations for the total cross
sections, by using DDR with the subtraction constant as a free fit
parameter and have shown that it affects the fit results at both
low and high energies; therefore, the subtraction constant can not be 
disregarded.
We have also shown that, for functions that are entire in the
logarithm of the energy (Taylor expansion), the derivative forms do not 
depend on the
parameter $\alpha$ \cite{am02}. More recently, Cudell, Martynov and
Selyugin have introduced a new representation for the DDR, intended for
low energies as well, and they also refer to the possible role of the
subtraction constant in the derivative form \cite{cms}.

Based on the above considerations, in the next sections we
investigate some formal and practical aspects related with the
replacement of the IDR by the DDR. Specifically we shall demonstrate
that, for entire functions in the logarithm of the energy,
the derivative forms do not depend on the $\alpha$ parameter and
shall discuss both the approximations involved and the important
practical and formal role of the subtraction
constant. Some of these results have already been presented elsewhere
\cite{am02}.

\section{Derivative dispersion relations} \label{sec:derivative}

In this section we first show that the subtraction constant is preserved when the 
IDR are replaced by the DDR and that, for functions entire in
the logarithm of the energy, the derivative forms do not depend on any 
additional free parameter. Next, based on the formulas displayed, we present
a detailed discussion on the conditions of validity of the 
different results mentioned
in the previous section, in connection with 
the assumptions involved. 
 
\subsection{Formal Derivation}

Let us consider the even amplitude, Eq. (\ref{eq:4}).
By defining
$s'=\e^{\xi'}$, $s=\e^{\xi}$ and
$g(\xi') = \im F_+(\e^{\xi'}) / \e^{\xi'}$, we
express

\begin{displaymath}
\re F_+(\e^{\xi}) - K=
\frac{2\e^{2\xi}}{\pi}P\!\!\int_{\ln s_0}^{+\infty}
\!\!\frac{g(\xi')\e^{\xi'}}{\e^{2\xi'}-\e^{2\xi}}\d\xi'
= \frac{\e^{\xi}}{\pi}P\!\!\int_{\ln s_0}^{+\infty}
\!\!\frac{g(\xi')}{\sinh(\xi'-\xi)}\d\xi'.
\end{displaymath}

Assuming that $g$ is an {\em analytical function of its argument\/}, 
we perform the expansion

\begin{displaymath}
g(\xi')=\sum_{n=0}^{\infty}
\left.\frac{\d^{n}}{\d\xi'^{n}}g(\xi')
\right|_{\xi'=\xi}\frac{(\xi'-\xi)^n}{n!}
\end{displaymath}
and then, the integration term by term in the above formula.
At the high energy limit,
we consider
the {\em essential approximation\/} $s_0 = 2m^2 \rightarrow 0$,
so that  $\ln s_0 \rightarrow -\infty$.
With these conditions, we obtain

\begin{displaymath}
\re F_+(\e^{\xi}) - K=
\frac{\e^{\xi}}{\pi}\sum_{n=0}^{\infty}\frac{g^{(n)}(\xi)}{n!}P
\!\!\int_{-\infty}^{+\infty}\!\!\frac{(\xi'-\xi)^{n}}{\sinh(\xi'-\xi)}
\d\xi'.
\end{displaymath}

Now, defining $y=\xi'-\xi$, the above formula may be put in the form

\begin{displaymath}
\re F_+(\e^{\xi}) - K= \e^{\xi}
\sum_{n=0}^{\infty}\frac{g^{(n)}(\xi)}{n!}I_n,
\end{displaymath}
where,

\begin{displaymath}
I_n=\frac{1}{\pi}P\!\!\int_{-\infty}^{+\infty}\!\!
\frac{y^{n}}{\sinh y}\d y.
\end{displaymath}

For $n$ even, $ I_n=0$ and for $n$ odd, we consider the integral

\begin{displaymath}
J(a)=\frac{1}{\pi}P\!\!\int_{-\infty}^{+\infty}\!\!
\frac{\e^{ay}}{\sinh y}\d y
= \tan \left(\frac{a\pi}{2} \right),
\end{displaymath}
so that,

\begin{displaymath}
I_n=\left.\frac{\d^n}{\d a^n}J(a)\right|_{a=0}=
\left.\frac{1}{\pi}P\!\!\int_{-\infty}^{+\infty}\!\!
\frac{\e^{ay}y^{n}}{\sinh y}\d y\right|_{a=0}.
\end{displaymath}

With this we have

\begin{eqnarray}
\re F_+(\e^{\xi}) - K&=&\e^{\xi}\sum_{n=0}^{\infty}\left.
\frac{1}{n!}\frac{\d^n}
{\d a^n}\tan\left(\frac{\pi a}{2}\right)
\right|_{a=0}\left.\frac{\d^n}{\d\xi'^{n}}g(\xi')\right|_{\xi'=\xi}
\nonumber\\
&=&\e^{\xi}\tan\left(\frac{\pi}{2}
\frac{\d}{\d\xi}\right)g(\xi)
\nonumber
\end{eqnarray}
and, therefore,

\begin{equation}
\frac{\re F_+(s)}{s}= \frac{K}{s} +
\tan\left[\frac{\pi}{2}
\frac{\d}{\d\ln s}\right]\frac{\im F_+(s)}{s},
\label{eq:10}
\end{equation}
where the series expansion is implicit in the tangent operator. With
analogous procedure for the odd relation we obtain

\begin{eqnarray}
\re F_-(s)=
\tan\left[\frac{\pi}{2}
\frac{\mathrm{d}}{\mathrm{d}\ln s}\right]
\im F_-(s),
\nonumber
\end{eqnarray}
or

\begin{equation}
\frac{\re F_-(s)}{s}=
\tan\left[\frac{\pi}{2}
\left(1 + \frac{\mathrm{d}}{\mathrm{d}\ln s}\right)\right]
\frac{\im F_-(s)}{s}.
\label{eq:11}
\end{equation}

We see that, with the exception of the subtraction constant, these
formulas do not depend on any additional free parameter, or, they
correspond to a ``particular" case of the BKS relations for $\alpha$
= 1. We stress that
the same result is obtained if, following BKS,
the derivation begins with an integration by parts without free
parameter, as can be easily verified. We shall return to this point 
in Section 3.2.

In practical uses the trigonometric operators are expressed by the
corresponding series. With the substitution in the odd case

\begin{displaymath}
\tan\left[\frac{\pi}{2}
\left(1 + \frac{\d}{\d\ln s}\right)\right]
\quad
\rightarrow
\quad
- \cot\left[\frac{\pi}{2}
\frac{\d}{\d\ln s}\right],
\end{displaymath}
and, by expanding the series {\em around the origin\/}, we obtain

\begin{eqnarray}
\frac{\re F_+(s)}{s} = \frac{K}{s} & +&
\bigg[ \frac{\pi}{2} \frac{\d}{\d\ln s} +
\frac{1}{3} \left(\frac{\pi}{2}\frac{\d}{\d \ln s}\right)^3
\nonumber \\
& + &
\frac{2}{5} \left(\frac{\pi}{2}\frac{\d}{\d \ln s}\right)^5 +
\cdots \bigg] \frac{\im F_{+}(s)}{s},
\label{eq:12}
\end{eqnarray}

\begin{eqnarray}
\frac{\re F_-(s)}{s}  =
&-& \frac{2}{\pi}\int \bigg\{ \bigg[
1 - \frac{1}{3} \left(\frac{\pi}{2}\frac{\d}{\d \ln s}\right)^2
\nonumber\\
&-&  \frac{1}{45} \left(\frac{\pi}{2}
\frac{\d}{\d \ln s}\right)^4
 - \cdots \bigg] \frac{\im F_{-}(s)}{s} \bigg\} \d \ln s,
\label{eq:13}
\end{eqnarray}
which is formally equivalent to the results introduced by Kang and
Nicolescu, Eqs. (\ref{eq:8}) and (\ref{eq:9}), except for the presence
of the subtraction constant.
These expansions have been used in the Ref. \cite{alm03}, that includes
power and logarithmic dependences in the parametrizations.

We have just demonstrated three formal and important results: (1) the
subtraction constant is preserved when the IDR are replaced by DDR and,
therefore, in principle, can not be disregarded;
(2) except for the subtraction constant, the DDR with entire functions
in the logarithm of the energy, do not depend on any
additional free parameter; (3) the only approximation involved in the
replacement concerns the lower limit in the IDR, namely,
$s_0 = 2m^2 \rightarrow 0$, which represents a high-energy 
approximation. We shall return to these points in Section 4,
when treating practical uses of the DDR.

\subsection{Discussion}

With focus on the DDR as expressed by Eqs. (6) to (13), let us
discuss here the different results mentioned in Section 2,
in connection with the corresponding conditions of validity and
assumptions involved. We shall roughly follow a chronological
approach (see also \cite{kfrev}).

The first derivative form appeared in the beginning of 1968 in the
context of the Regge Theory. By investigating the cuts associated with
the Pomeron in the high energy limit ($s \rightarrow \infty$) and at
low momentum transfers, Gribov and Migdal introduced a derivative
representation for the Watson-Sommerfeld integral, which, mathematically,
corresponds to the first term in the expansion (12) (without the
subtraction constant), namely

\begin{eqnarray}
\frac{\re F_+(s,t)}{s} =
\bigg[ \frac{\pi}{2} \frac{\partial}{\partial\ln s} 
\bigg] \frac{\im F_{+}(s,t)}{s}.
\nonumber
\end{eqnarray}

In the beginning of the seventies, Bronzan \cite{bronzan68}, followed
by Jackson \cite{jack73}, introduced the first form involving the
tangent operator, Eqs. (10) and (11), independent of the
parameter $\alpha$. The arguments by Jackson were based on
the formal representation of the Taylor series by an exponential
operator,

\begin{eqnarray}
f(z + \lambda) =
\exp \bigg\{ \lambda\frac{d}{dz} \bigg\} f(z),
\nonumber
\end{eqnarray}
an on the high energy limit, in order to avoid branch points and cuts
(see \cite{jack73} for details). Therefore, the basic assumptions concern entire
functions and the high energies.

The first direct connection between IDR and DDR appeared in 1974 in the
work by BKS \cite{bks}. Starting from the singly subtracted IDR, Eq. (4)
(neglecting the subtraction constant), the authors mention an integration by
parts, which leads to an expression containing the parameter $\alpha$.
That may be obtained by multiplying and dividing Eq. (4) 
by $s^{\alpha}$ and then
integrating by parts. In addition to the high-energy approximation,
represented by the lower limit $s_0 \rightarrow 0$, a Taylor expansion of
$\im F_+/s^{\alpha}$ is also assumed, so that the series may be integrated
term by term (uniform convergence), leading to Eqs. (6) and (7). We should
note that the expansion considered by BKS in \cite{bks} is around the
point $\pi(\alpha - 1)/2$ for the even amplitude and $\pi \alpha/2$ in
the odd case, which leads to the previously mentioned singularity for 
$\alpha = 1$.
Differently, Eqs. (8), (9) and (12), (13) and all the other cases treated 
here and discussed
in what follows, refer to expansions around the origin.

In the same year, Kang and Nicolescu introduced the series expansion
(8) and (9),
which, according to the authors, can be derived ``from the analyticity
relation given by the Sommerfeld-Watson-Regge representation" and are
asymptotic relations \cite{kn}. The point here was to treat the possibility
that the amplitude could include not only simple poles in the complex
angular momentum plane but, in particular, logarithm dependences in the
odd amplitude, representing the Odderon \cite{odd}. The association of this
concept with a complicated singularity at $J$ =1, demanded a specific
contour for the Watson-Sommerfeld-Regge transformation, which could be
translated by the need of subtraction in the dispersion relation \cite{kn}.
That, in turn, could be represented by the derivative factor,
$(\pi/2) d/dln s$, in the odd case, Eq. (9) \cite{pc}.
As we have shown, the DDR by Kang-Nicolescu and that by BKS have the same
mathematical structure, once the cotangent operator (Section 3.1) could be
well defined for the specific class of functions involved. We shall return 
to this important point in Section 5.

At that time, the derivative approach was strongly criticized by several authors
\cite{crit1,crit2,crit3,crit4}, in the sense that it had no practical interest, 
for instance, because ``the mathematical condition for the convergence of the
series excludes all cases of physical interest" \cite{crit1}. However,
it should be noted that some of these criticisms were connected with the
physical situation at that time, some directed to attempts to extend the method
to low energies (for example, as in \cite{suketal}) and/or to some possible
excessive optimism about the derivative approach \cite{sidhu}. As effective
contributions from all these criticisms we may mention the following results,
which refer always to the even amplitude.

Eichmann and Dronkers treated
the case corresponding to $\alpha$ = 1, showing that Eq. (10) is valid only 
for some class of 
entire functions of $\ln s$ \cite{crit1}. Specifically, with our notation of 
Sec. $3.1$, they considered the series expansion of

\begin{eqnarray}
\tan \bigg[ \frac{\pi}{2} \frac{d}{d \xi} 
\bigg] g(\xi)
\nonumber
\end{eqnarray}
and have proved that, if this series converges, then $g(\xi)$ must be an entire 
function of $\xi$ and must satisfy $|g(\xi)| \leq C\exp\{|\xi|\}$, $C$ a
constant. Conversely,
if $g(\xi)$ is an entire function of $\xi$ and satisfies
$|g(\xi)| \leq C\exp\{(1 - \epsilon)|\xi|\}$, $\epsilon > 0$, than the tangent 
series converges and may be represented by the integral relation (4), with $s_0 = 0$.
Basically, the proof consisted in expanding $g(\xi)$ in Taylor series and
compare the result with that obtained by performing the same expansion of
$g(\xi)$ in the integral (4). They have also shown that the method does
not apply in the resonance region. 

Heidrich and Kazes considered the parameter $\alpha$ and
showed that the convergence of the series demands that
$\alpha$ must lie between 0 and 2 \cite{crit2}. As before, it is assumed that
$\im F_{+}(s) / s^{\alpha}$ may be expanded in Taylor series and that, in turn,
integrated term by term. That leads to the evaluation of an integral
depending on $y = \xi - \xi'$, in the form

\begin{eqnarray}
\frac{1}{\pi} \ln\coth\frac{1}{2}|y|\:\: e^{(\alpha -1)y}y^{n}
{\large \mid}_{-\infty}^{+\infty}
+ \frac{1}{\pi}\int_{-\infty}^{+\infty}dy \frac{e^{(\alpha -1)y}}{\sinh y}y^{n},
\nonumber
\end{eqnarray}
so that the first term is finite only for
0 $ < \alpha < $ 2. They have also shown that Eq. (6) is violated in the
energy intervals between two branch points on the cut. It may be noted
that a criterion of convergence based on the convergence radius of the
Taylor expansion was later shown to be incorrect by 
Fischer and Kol\'a\v{r} \cite{kf}.

From 1976 to 1987, Fischer and Kol\'a\v{r}, in a series of seminal papers
\cite{kf}, developed a rigorous study of the derivative approach in connection 
with general principles and theorems from axiomatic quantum field theory \cite{ft}.
The authors introduced classes of functions satisfying, among other properties,
polynomial bound and the Froissart-Martin bound, and corresponding to a wider class
of functions than that represented by entire functions in the logarithm of the
energy. The analysis was focused on the region of asymptotic energies and the
main formal result was that the derivative relations are valid if the tangent 
series is replaced by its first term. Although that demands the existence 
of the high-energy limits of certain physical quantities \cite{kf},
the class of functions involved includes the majority of functions of
interest in physical applications. 
For our purposes, we recall some results
obtained by the authors about the role and range of the parameter $\alpha$.
For $x = \ln s$, let $F(x)$ be a function belonging to the above mentioned class of 
functions 
\cite{kf}. For $\alpha$ = 1, the entire function $\mathcal{F}(z)$
which extends the function $F(x)$ to the complex $z$ plane, obeys the bound

\begin{eqnarray}
|\mathcal{F}(z)| \leq \epsilon e^{|z|} + C(\epsilon),
\nonumber
\end{eqnarray}
for every $\epsilon >$ 0, where 
$C(\epsilon)$ is a constant that depends on $\epsilon$. On the other
hand, for $\alpha \not=$ 1, if the series

\begin{eqnarray}
\tan \bigg[ \frac{\pi}{2} 
(\alpha - 1 + \frac{d}{dx}) 
\bigg] F(x)
\nonumber
\end{eqnarray}
converges on some interval, then $F(x)$ can be extended to an entire 
function for any $\alpha$ and in this case, the entire function obeys

\begin{eqnarray}
|\mathcal{F}(x)| \leq \epsilon e^{|x| - (\alpha - 1) \re x} + 
C(\epsilon)\epsilon e^{- (\alpha - 1) \re x}.
\nonumber
\end{eqnarray}
Therefore, the bound changes, unless $x$ lies on the imaginary
axis \cite{kf}.
To our knowledge that is the only result in the literature that
``quantify" the role of the parameter $\alpha$. We shall return to this 
point in what follows.
We also recall that differently from all the other works discussed here,
Fischer and Kol\'a\v{r} make explicit reference to the fact that the
tangent series represents the difference between $\re F_{+}/s$ and
a constant associated with the subtraction. Moreover, they
also discuss the contribution from the integral with zero and $s_0$ as
lower and upper limits, respectively 
(the high-energy approximation). They have also
shown that for entire functions associated with fits, this contribution
is divergent unless $\im F(0)$ = 0 (a result also presented in
\cite{crit2}).

It should be also mentioned that Block and Cahn obtained a 
result for the DDR that
does not depend on $\alpha$, but under the assumption that the
differential operator $ Z = \d/\d\ln s$ follows the condition
$|Z| < 1$ \cite{bc}.

We conclude this discussion with the following comments. Concerning the 
range and validity of the parameter $\alpha$, we have shown that 
the different statements, briefly 
quoted in Section 2, are not controversial since they are consequences of
the different assumptions or classes of functions considered. As shown by
Fischer and Kol\'a\v{r}, for functions that are entire in $\ln s$, the DDR are 
continuation of the IDR for all values of the parameter $\alpha$.
However, based on the above bounds, established for 
$\alpha$ = 1 and $\alpha \not=$ 1, we can not devise any practical 
advantage in using $\alpha \not=$ 1, as stated before by
Fischer and Kol\'a\v{r}. In this sense, we understand that the introduction of
the parameter $\alpha$ by BKS is only an unnecessary complication.

With the exception of the analysis by
Fischer and Kol\'a\v{r}, neither of the works discussed in this section
make any reference to the subtraction constant nor treat explicitly the
contribution from the high-energy approximation represented 
by the assumption $s_0 \rightarrow$ 0.
In the following section we investigate the practical role of
the subtraction constant and the influence of the lower limit
of the IDR for a class of functions that are entire in $\ln s$.

\section{Tests with Pomeron-Reggeon parametrizations} \label{sec:tests}

In order to quantitatively investigate the practical applicability of
the DDR in substitution to IDR,
we consider, as a framework, some standard parametrizations for the
total cross section and present a detailed study on simultaneous fits
to $\sigma_{\mathrm{tot}}(s)$ and $\rho(s)$. We start this Section with some comments
about our choices concerning the parametrizations, ensemble of
experimental data, energy cutoffs and an outline on the strategies
to be used in this study. We then present the fit results, followed
by a discussion on the role of the subtraction constant and of the
high-energy approximation associated with the lower integral limit.

\subsection{Basic comments}

Presently, phenomenological analyses on the forward quantities
$\sigma_{\mathrm{tot}}(s)$ and $\rho(s)$, at high energies,
 are based on two kinds of dependences on
the energy, namely power and power logarithmic functions and, therefore,
entire functions in $\ln s$. In the context of the Regge
phenomenology they are associated with simple-pole Pomeron and secondary reggeons
($s^{\pm \gamma}$, 0 $< \gamma < $ 1), double-pole ($\ln s$) and triple-pole
($\ln^2 s$) contributions \cite{competework} . In order to treat in detail 
the applicability of the
DDR and IDR by means of a complete example, we shall consider here,
as a framework, only parametrizations based on simple-pole Pomeron and
secondary reggeons, leaving the other cases for a forthcoming work.

Although that Pomeron contribution, represented by the $s^{\epsilon}$
dependence, with $\epsilon \approx 0.081$ (Section 4.2), eventually violates the 
Froissart-Martin bound, we shall focus here only on its mathematical
character, as representing a class of functions entire in $\ln s$ and,
moreover, that can provide a good description of the experimental data 
at the energies presently available. It should be also noted that other
forms of DDR do not require the above bound, as demonstrated in
\cite{kf}.

Since all the Regge phenomenology is intended for the region of high 
energies \cite{ddln} and, as mentioned previously, for particle-particle
and particle-antiparticle (crossing) the $pp$ and $\bar{p}p$
scattering correspond to the highest energy values
with available data, we shall concentrate here only on these two
processes. We shall return to this point in Section 4.4.
We make use of the data sets on $\sigma_{\mathrm{tot}}$ and $\rho$
analyzed and compiled by the Particle Data Group \cite{pdg}.
The statistic and systematic errors have been added in 
quadrature.

Since we are interested in the practical role of the subtraction
constant as a free fit parameter, it is necessary to test both the different
cutoffs in the energy and the effect of the number of  free
fit parameters involved. To treat the former case, we shall consider in all
the analysis two lower energy cuts: $\sqrt s_{\mathrm{min}} =$ 5 GeV and
$\sqrt s_{\mathrm{min}} =$ 10 GeV. In the later case,
in order to get quantitative information on the sensitivity of
the subtraction constant as a free fit parameter, 
we first consider an ``economical" parametrization represented by
two reggeon exchanges
(the Donnachie-Landshoff model) and after,
an extended model with three reggeon exchanges. In the Regge context 
that means to consider {\em degenerate\/} and {\em non-degenerate\/} 
higher {\em meson trajectories\/},
respectively.

Our aim in this section is to use these parametrizations in order
to discuss the applicability of both the IDR, either with $s_0 = 2m^2$ or
$s_0 = 0$, and the DDR. Specifically, we want to establish
how the results with the DDR deviate from those with the IDR and in
which circumstances both approaches lead to the same results.
Summarizing, our strategy shall be to consider simultaneous fits 
to $\sigma_{\mathrm{tot}}$ 
and $\rho$ data from $pp$ and $\bar{p}p$ scattering, with the following variants:
(i) parametrizations with degenerate and non-degenerate trajectories;
(ii) lower energy cuts  at $\sqrt s_{\mathrm{min}} =$ 5 and $\sqrt s_{\mathrm{min}} =$ 10 GeV;
(iii) subtraction constant: $K$ = 0 and $K$ as a free fit parameter;
(iv) DDR and IDR with both $s_0 = 2m^2$ and with $s_0 =0$.

In what follows, we consider the approximation
$s = 4(k^2 + m^2) \sim 4 k^2$, so that the optical theorem,
Eq. (\ref{eq:2}) reads

\begin{equation}
\sigma_{\mathrm{tot}}(s) = \frac{\im F(s,t=0)}{s}.
\label{eq:14}
\end{equation}

\subsection{Degenerate meson trajectories}

The forward effective Regge amplitude introduced by {\em Donnachie and Landshoff}
(DL)
has two contributions, one from a single Pomeron and the other from
secondary Reggeons exchanges \cite{dl}.
The parametrization assumes degeneracies between the secondary reggeons, imposing a common 
intercept for the $C=+1$ ($a_2, f_2$) and the $C=-1$ ($\omega,\rho$)
and is given by

\begin{equation}
\sigma_{\mathrm{tot}}^{pp} (s) = X s^{\epsilon} + Y s^{- \eta},
\label{eq:15}
\end{equation}

\begin{equation}
\sigma_{\mathrm{tot}}^{p\bar{p}} (s) = X s^{\epsilon} + Z s^{- \eta}.
\label{eq:16}
\end{equation}
Here, $\epsilon=\alpha_{\mathbb{P}}(0)-1$ and
$\eta= 1 - \alpha_{\mathbb{R}}(0)$, where $\alpha_{\mathbb{P}}(0)$ and
$\alpha_{\mathbb{R}}(0)$ are the
Pomeron and Reggeon intercepts, respectively.
From analysis of $pp$ and $\bar{p}p$ in the interval
$10\:\:{\mathrm{GeV}} \leq \sqrt{s} \leq 546\:\:{\mathrm{GeV}}$,
DL
obtained the following values for the free parameters:
$X = 21.7$ mb, $Y = 56.08$ mb, $Z = 98.39$ mb, $\epsilon = 0.0808$,
$\eta = 0.4525$.

In what follows, we first present the analytical connections between 
$\sigma_{\mathrm{tot}}$ and $\rho$
with the DL parametrization, using both the IDR
(with $s_0 = 2m^2$ or $s_0 = 0$) and the DDR and then, the results
of global (simultaneous) fits to
$\sigma_{\mathrm{tot}}$ and $\rho$, with $K=0$ or $K$ as a free fit 
parameter and energy cutoffs at 5 and 10 GeV.

\subsubsection{Analytical results}

The crossing even and odd amplitudes in Eq. (\ref{eq:1}) are obtained by
using the parametrizations (15) and (16) and the 
optical theorem as given
by Eq. (14). In this case, $\im F_{+/-}(s) \propto s^{\gamma}$,
with $-1 < \gamma < 1$ and the integrals in (\ref{eq:4}) and 
(\ref{eq:5}) may be analytically evaluated. 
For $s_0=0$ we obtain,

\begin{displaymath}
P\!\int_{0}^{+\infty} \d s'
\frac{s'^{\gamma} }{(s'^{2}-s^{2})} = \frac{\pi}{2} s^{ \gamma - 1}
\tan\frac{\pi \gamma}{2}.
\end{displaymath}

For $s_0=2m^2$ fixed, the change of variables $s' = y + s_0$
and integration in $y$,
lead to the result 

\begin{eqnarray}
P\int^{+\infty}_{s_0}\!\!\frac{ s'^{\gamma}}{s'^2-s^2}\d s' =
\frac{\pi}{2} s^{ \gamma - 1}\tan\frac{\pi \gamma}{2} + L(s),
\nonumber
\end{eqnarray}
where, from the above formulas, $L(s)$ represents the contribution below the 
lower integration limit, 

\begin{displaymath}
L(s) = -P\!\int_{0}^{s_0} \d s'
\frac{s'^{\gamma} }{(s'^{2}-s^{2})} = \frac{\pi}{2} s^{ \gamma - 1}
\tan\frac{\pi \gamma}{2},
\end{displaymath}
and can be expressed
either interms of hypergeometric functions or the corresponding series
expansion in inverse powers of $s$ \cite{tables}

\begin{eqnarray}
L(s)  &=& \frac{s{_0}^{\gamma}}{2\gamma s}\left[
_{2}\mathrm{F}_{1}(1,\gamma;1+\gamma;s_0/s)-\:
_{2}\mathrm{F}_{1}(1,\gamma;1+\gamma;-s_0/s)\right] \nonumber \\
&=& \frac{s{_0}^{\gamma}}{s}
\sum_{j=0}^{\infty}
\frac{1}{2j+1+\gamma}\left(\frac{s_0}{s}\right)^{2j+1}. \nonumber
\end{eqnarray}

With these expressions, and the DL parametrization, the connections
between $\sigma_{\mathrm{tot}}(s)$ and $\rho(s)$ may be
analytically determined.
For $s_0=0$ we obtain

\begin{eqnarray}
\rho(s) \sigma_{\mathrm{tot}}(s) & = & \frac{K}{s}
\pm \frac{(Y-Z)}{2}s^{-\eta}\cot \left(\eta\frac{\pi}{2}\right)
+ Xs^{\epsilon}\tan \left(\epsilon\frac{\pi}{2}\right)
\nonumber\\
&-&
\frac{(Y+Z)}{2}s^{-\eta}\tan \left(\eta\frac{\pi}{2}\right),
\label{eq:17}
\end{eqnarray}
and for fixed $s_0=2m^2$,

\begin{eqnarray}
\rho(s) \sigma_{\mathrm{tot}}(s) & = & \frac{K}{s}
\pm \frac{(Y-Z)}{2}s^{-\eta}\cot \left(\eta\frac{\pi}{2}\right)
+ Xs^{\epsilon}\tan \left(\epsilon\frac{\pi}{2}\right)
\nonumber\\
&-&
\frac{(Y+Z)}{2}s^{-\eta}\tan \left(\eta\frac{\pi}{2}\right)
+ 
\frac{1}{\pi}
\sum_{j=0}^{\infty}
\bigg(
\pm  \frac{(Y-Z)s{_0}^{1-\eta}}{s(2j+2-\eta)}
\nonumber\\
&+& \frac{2Xs{_0}^\epsilon}{2j+1+\epsilon}
+\frac{(Y+Z)s{_0}^{-\eta}}{2j+1-\eta}
\bigg)\left(\frac{s_0}{s}\right)^{2j+1},
\label{eq:18}
\end{eqnarray}
where the signs $\pm$ apply for $pp$ ($+$) and $\bar{p}p$ ($-$)
scattering and, as before, the power series is associated
with the contribution below the lower integration
limit.

The results with the DDR, as given by Eqs. (\ref{eq:10})
and (\ref{eq:11}) (or (\ref{eq:12}) and (\ref{eq:13})),
are exactly the same as that obtained by means of the IDR with 
$s_0 = 0$,  namely, Eq. (\ref{eq:17}).
This is in agreement with our conclusion at the end of the  Section $3.1$:
the only formal approximation involved in going from IDR to DDR
is to take $s_0 = 2m^2 \rightarrow 0$.

Once we have established the analytical equivalence of the real parts
obtained with both DDR and IDR for $s_0 = 0$, in what folllows we shall
consider only the results provided by DDR, Eq. (17), and by
IDR for $s_0 = 2m^2$, Eq. (18).

\subsubsection{Global Fits}

\begin{figure}[ht]
\includegraphics[width=7cm,height=7cm]{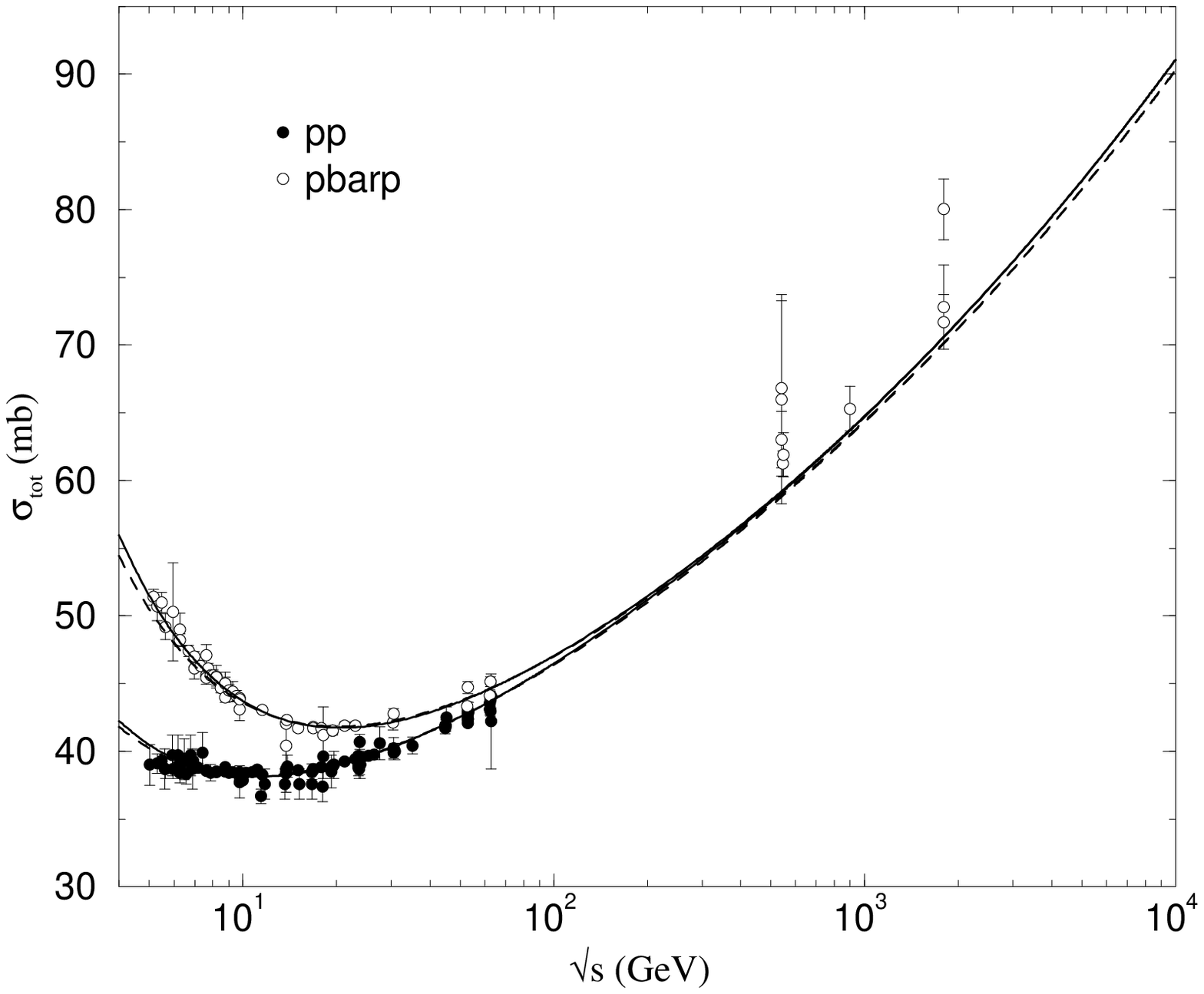}
\includegraphics[width=7cm,height=7cm]{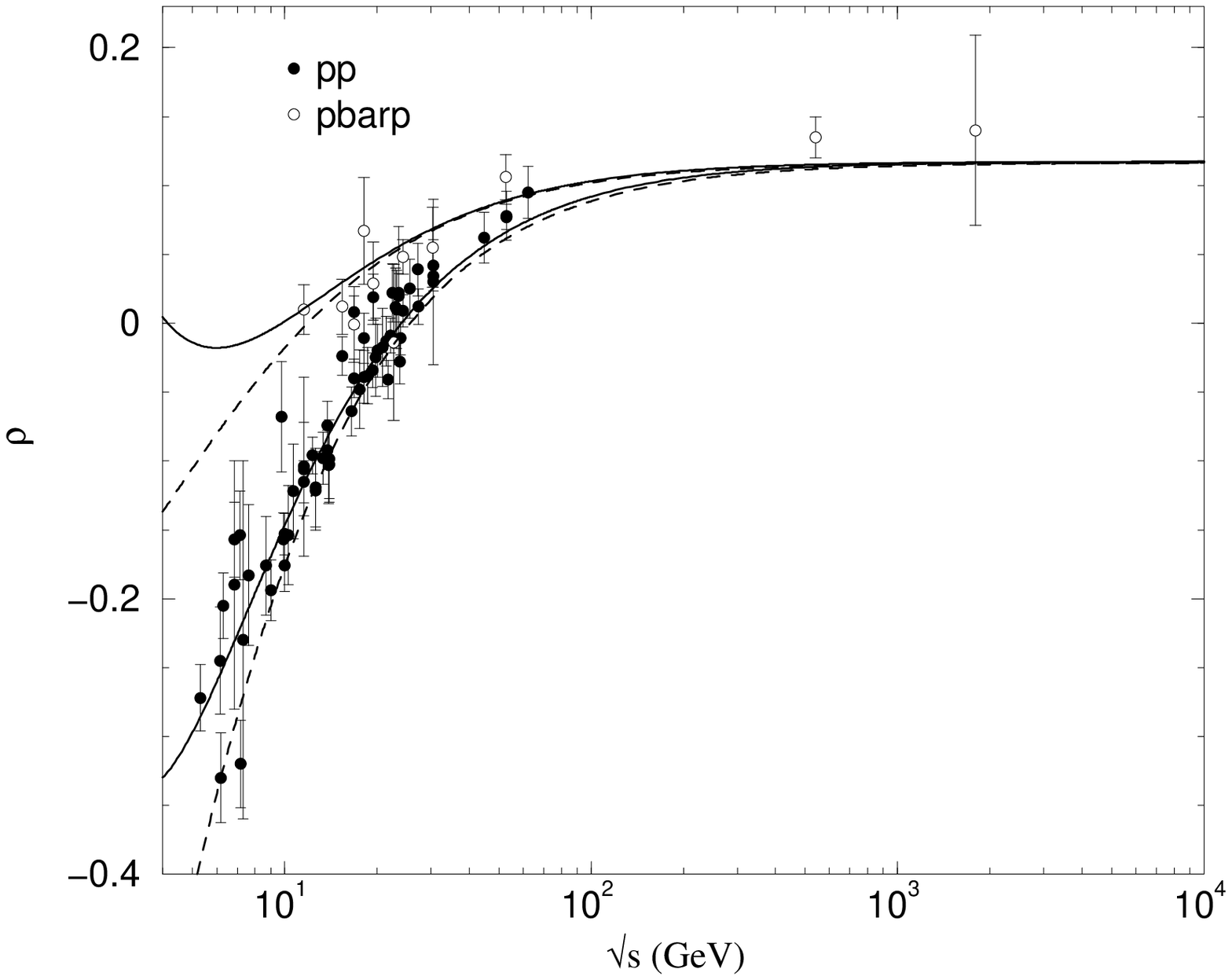}
\caption{Simultaneous fit to $\sigma_{\mathrm{tot}}$ and $\rho$
through the DL parametrization, $\sqrt s_{\mathrm{min}} =$ 5 GeV,
assuming $K=0$ and using either the IDR with $s_0=2m^2$ (solid) or the
DDR (dashed).} \label{fig:1}
\end{figure}

\begin{figure}[ht]
\includegraphics[width=7cm,height=7cm]{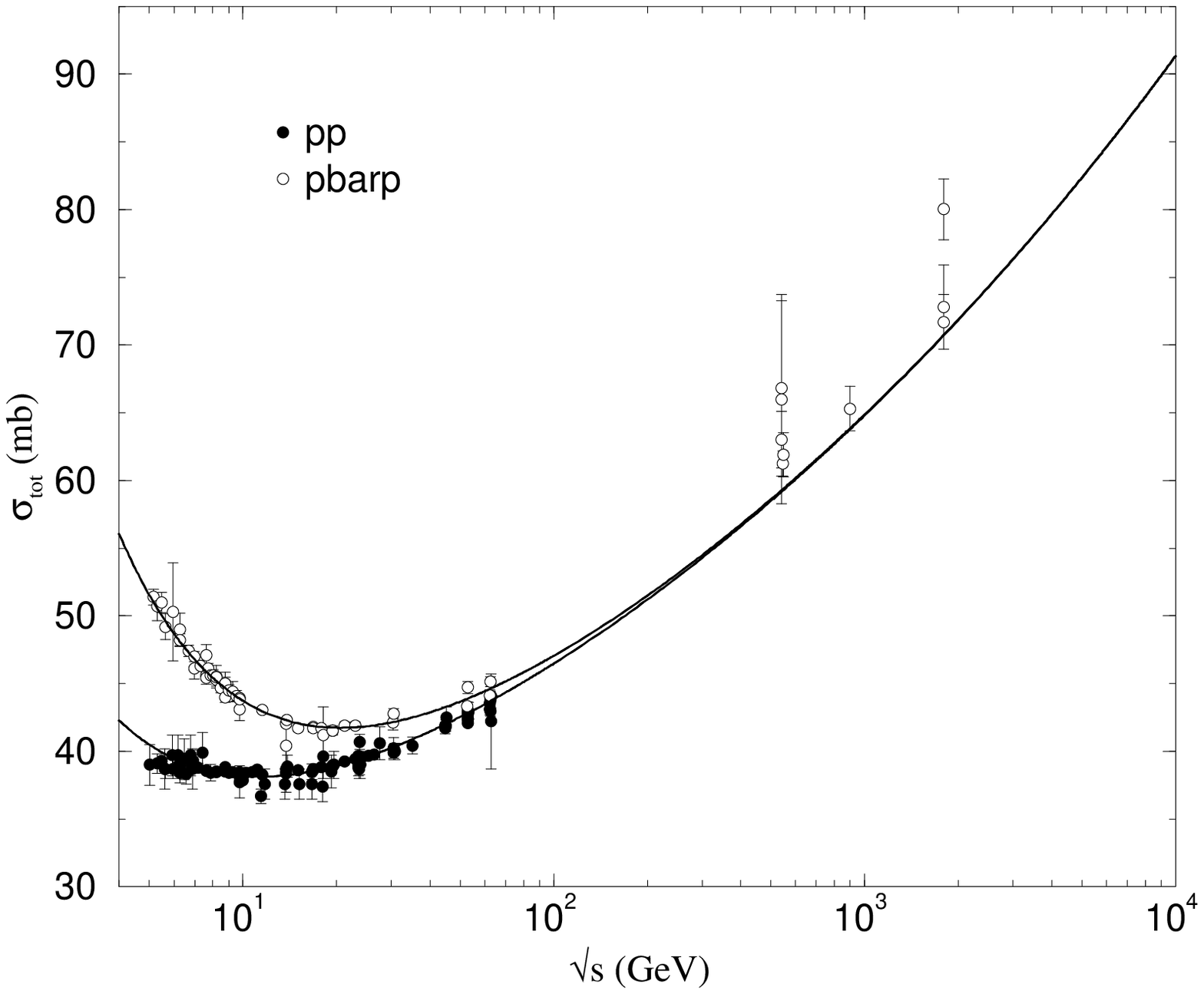}
\includegraphics[width=7cm,height=7cm]{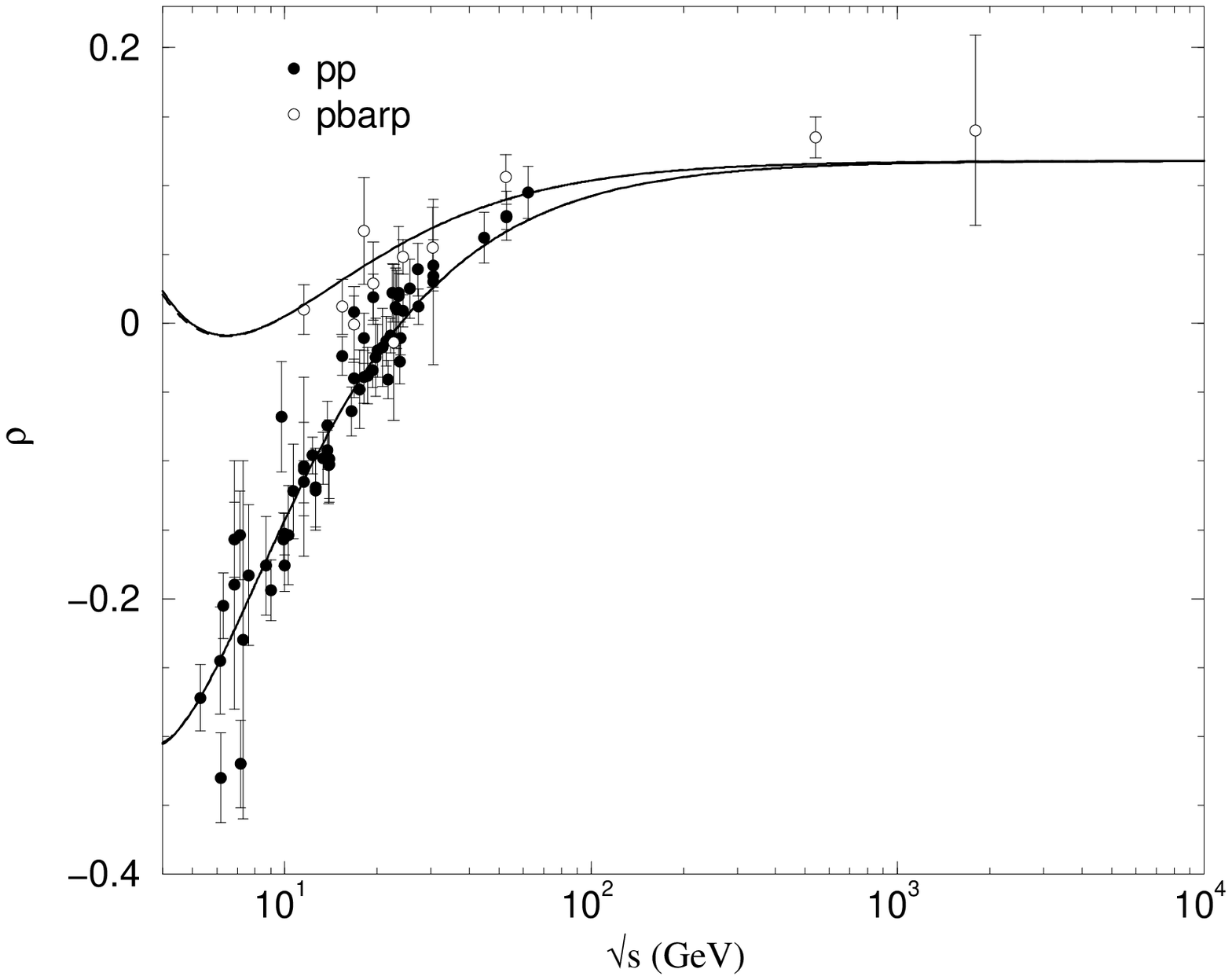}
\caption{Simultaneous fit to $\sigma_{\mathrm{tot}}$ and $\rho$ 
through the DL parametrization, $\sqrt s_{\mathrm{min}} =$ 5 GeV, with $K$
as free parameter and using either the IDR with $s_0=2m^2$ (solid)
or the DDR (dashed): both curves coincide.} \label{fig:2}
\end{figure}

\begin{figure}[ht]
\includegraphics[width=7cm,height=7cm]{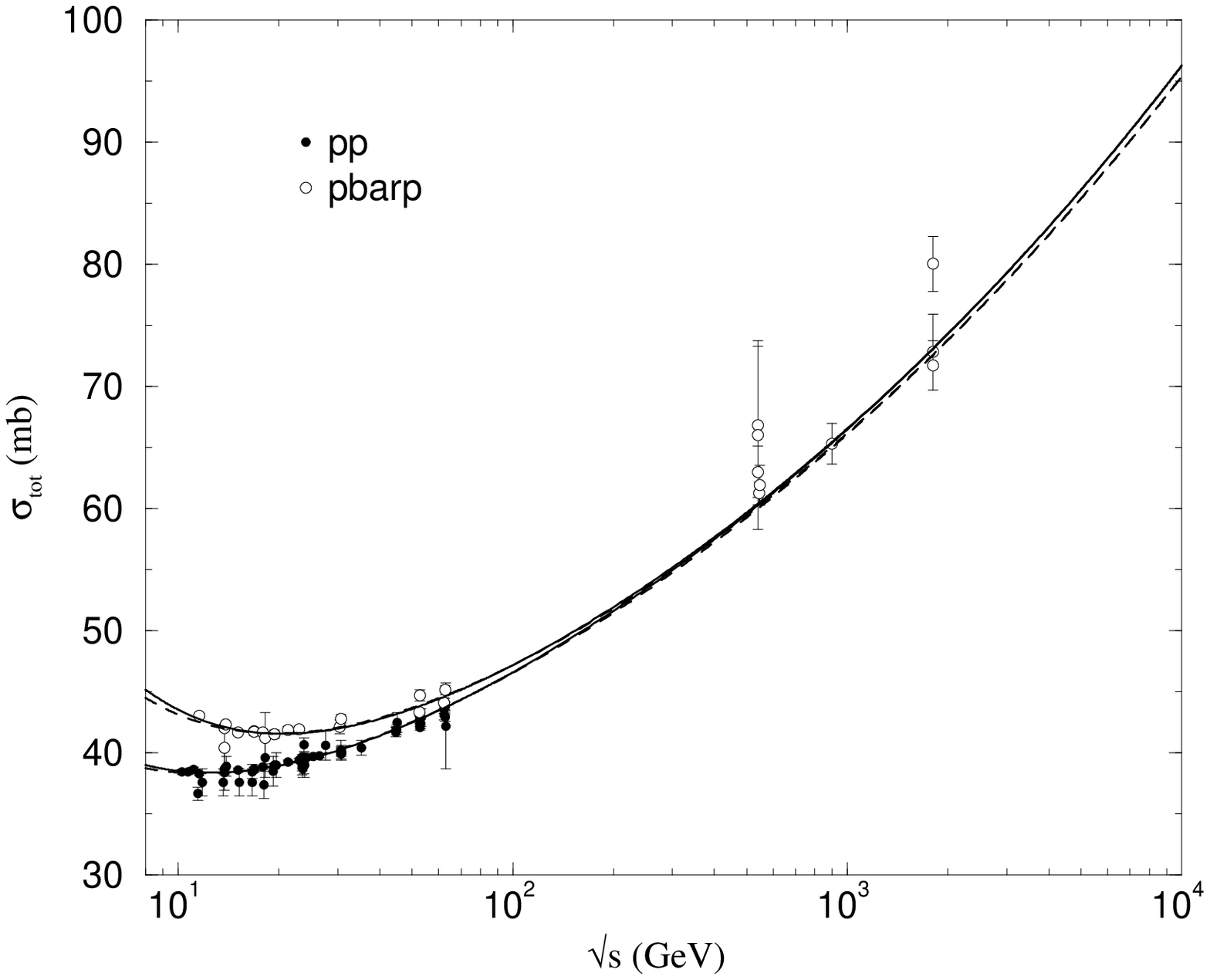}
\includegraphics[width=7cm,height=7cm]{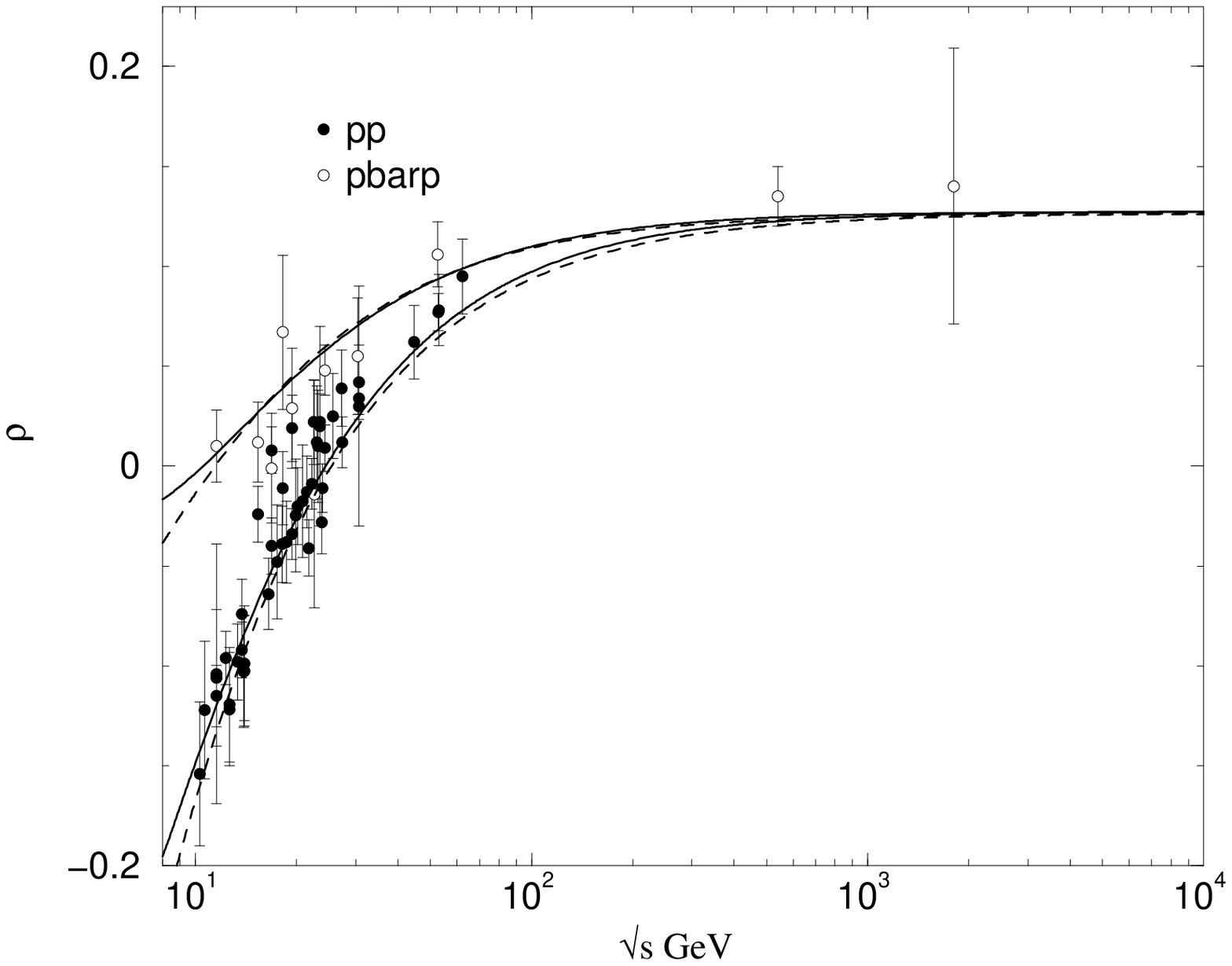}
\caption{Simultaneous fit to $\sigma_{\mathrm{tot}}$ and $\rho$
through the DL parametrization, $\sqrt s_{\mathrm{min}} =$ 10 GeV,
assuming $K=0$ and using either the IDR with $s_0=2m^2$ (solid) or the
DDR (dashed).} \label{fig:3}
\end{figure}

\begin{figure}[ht]
\includegraphics[width=7cm,height=7cm]{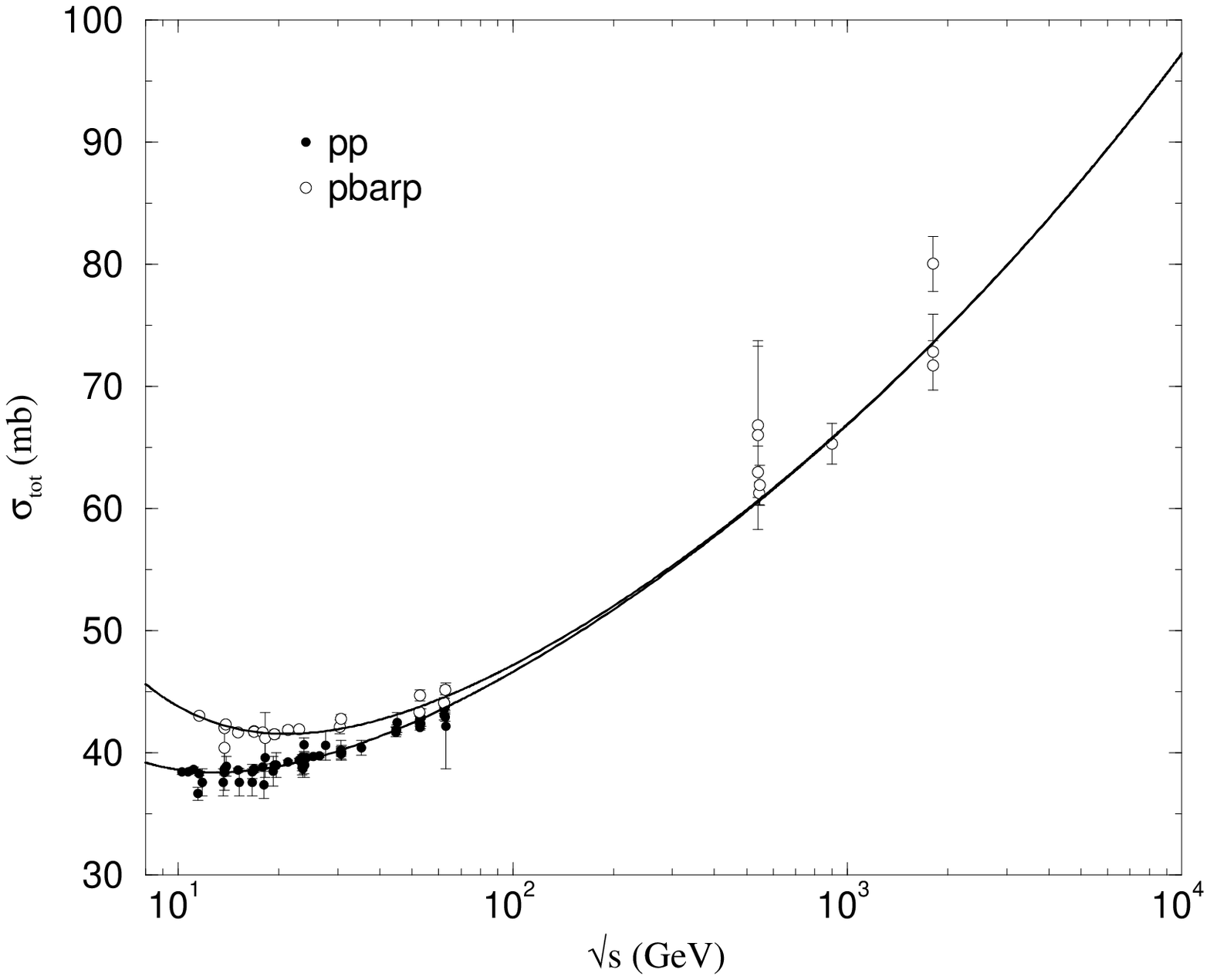}
\includegraphics[width=7cm,height=7cm]{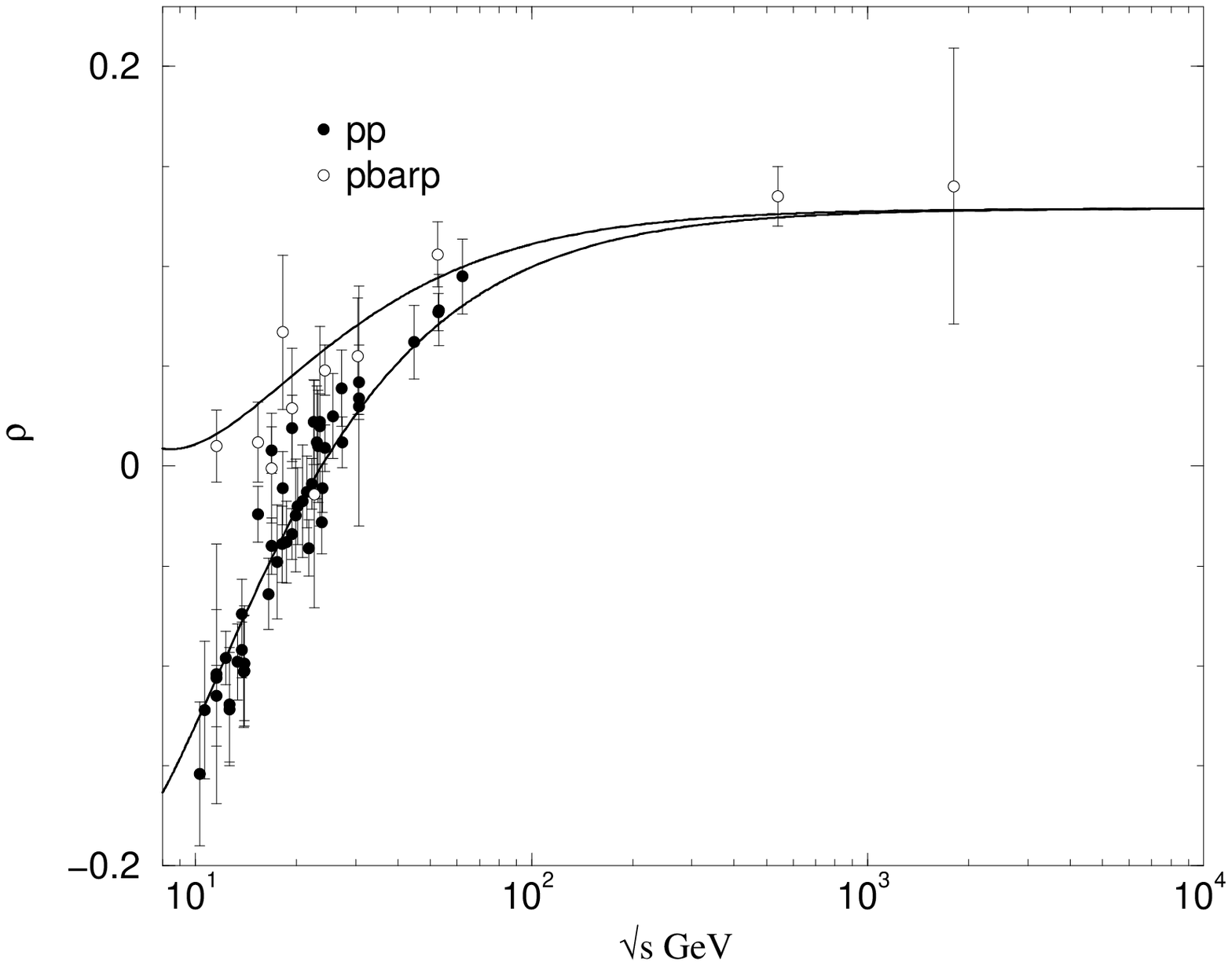}
\caption{Simultaneous fit to $\sigma_{\mathrm{tot}}$ and $\rho$ 
through the DL parametrization, $\sqrt s_{\mathrm{min}} =$ 10 GeV, with $K$
as free parameter and using either the IDR with $s_0=2m^2$ (solid)
or the DDR (dashed): both curves coincide.} \label{fig:4}
\end{figure}

We have developed global (simultaneous) fits to $\sigma_{\mathrm{tot}}$ and
$\rho$ with the DL param\-etri\-za\-tion, using either the IDR
with $s_0=2m^2$, or the DDR,
considering either $K=0$ or
$K$ as a free fit parameter and for energy cuts at
5 and 10 GeV. The fits have been performed with the
program CERN Minuit and the errors in the fit parameters correspond 
to an increase of the $\chi^2$ by one unit.
The numerical results and statistical information
from all the four cases analyzed with energy cut at 5 GeV are displayed in 
Table \ref{tab:1},
and the curves, together with the experimental data, are shown 
in Figs. \ref{fig:1} and \ref{fig:2}, for $K=0$ and $K$ as a free fit 
parameter, respectively. The corresponding results for the 
energy cut at 10 GeV are presented
in Table \ref{tab:2} and Figs. \ref{fig:3} and \ref{fig:4}.
We shall discuss these results in Section 4.4 together with those 
obtained by means of an extended parametrization, characterized by the
non-degenerate trajectories. 

\begin{table}[h]
\begin{center}
\caption{Simultaneous fits to $\sigma_{\mathrm{tot}}$ and $\rho$
through
the DL parametrization, $\sqrt s_{\mathrm{min}} =$ 5 GeV (238 data points), 
either with $K$ as a free
parameter or $K=0$ and using IDR with lower limit $s_0=2m^2$ and DDR.}
\label{tab:1}
\begin{tabular}{ccccc}
\hline
&\multicolumn{2}{c}{IDR with $s_0=2m^2$}&\multicolumn{2}{c}{DDR}\\
\hline
 &    $K$ free         & $K=0$               &     $K$ free          & $K=0$                    \\
\hline
X (mb)       & 23.08 $\pm$ 0.28    & 23.14  $\pm$ 0.28   & 23.08   $\pm$ 0.28   & 23.17   $\pm$ 0.28  \\
Y (mb)       & 54.65 $\pm$ 1.3     & 54.4   $\pm$ 1.3    & 54.66   $\pm$ 1.3    & 48.8    $\pm$ 1.0   \\
Z (mb)       & 108.8  $\pm$ 3.5    & 108.5  $\pm$ 3.5    & 108.8   $\pm$ 3.6    & 95.0    $\pm$ 2.4    \\
$\epsilon$   & 0.0747 $\pm$ 0.0014 & 0.0743 $\pm$ 0.0014 & 0.0747  $\pm$ 0.0014 & 0.0738  $\pm$0.0014   \\
$\eta$       & 0.494  $\pm$ 0.010  & 0.4946 $\pm$ 0.0099 & 0.494   $\pm$ 0.010  & 0.4673  $\pm$ 0.0079  \\
$K$          & 17     $\pm$ 13     & 0                   & 179     $\pm$ 15     &0                      \\
$\chi^2/F$ & 1.39                & 1.39                & 1.39                 &2.01               \\
\hline
\end{tabular}
\end{center}
\end{table}


\begin{table}[h]
\begin{center}
\caption{Simultaneous fits to $\sigma_{\mathrm{tot}}$ and $\rho$ 
through
the DL parametrization, $\sqrt s_{\mathrm{min}} =$ 10 GeV (154 data points), 
either with $K$ as a free
parameter or $K=0$ and using IDR with lower limit $s_0=2m^2$ and DDR.}
\label{tab:2}
\begin{tabular}{ccccc}
\hline
&\multicolumn{2}{c}{IDR with $s_0=2m^2$}&\multicolumn{2}{c}{DDR}\\
\hline
 &    $K$ free         & $K=0$               &     $K$ free          & $K=0$                    \\
\hline
X (mb)       & 21.62  $\pm$ 0.38   &21.78  $\pm$ 0.37   & 21.62  $\pm$  0.38   &21.83    $\pm$  0.38  \\
Y (mb)       & 65.7   $\pm$ 4.5    &59.5   $\pm$ 3.2    & 65.7   $\pm$  4.5    &50.6     $\pm$   2.2   \\
Z (mb)       & 113.4  $\pm$ 9.4    &102.5  $\pm$ 7.0    & 113.4  $\pm$  9.4    &85.7     $\pm$   4.4   \\
$\epsilon$   & 0.0816 $\pm$ 0.0018 &0.0806 $\pm$ 0.0018 & 0.0816 $\pm$  0.0018 &0.0800   $\pm$   0.0018   \\
$\eta$       & 0.482  $\pm$ 0.019  &0.467  $\pm$ 0.017  & 0.482  $\pm$  0.019  &0.435    $\pm$   0.014 \\
$K$          & 116    $\pm$ 36     &0                   & 287    $\pm$  44     &0\\
$\chi^2/F$ & 1.17                &1.24                & 1.17                 &1.50 \\
\hline
\end{tabular}
\end{center}
\end{table}

\vspace{1.0cm}

\subsection{Non-degenerate meson trajectories}

Analyses, treating global fits to
$\sigma_{\mathrm{tot}}$ and $\rho$, have indicated that the best results are 
obtained with 
non-degenerate meson trajectories \cite{nondege}. In this case the forward
scattering amplitude is decomposed into three reggeon exchanges,
$
F(s) = F_{\tt I\!P}(s) + F_{a_2/f_2}(s) + 
\tau F_{\omega/\rho}(s), 
$
where the first term represents the exchange of a single Pomeron, the other
two the secondary Reggeons and $\tau = + 1$ ($- 1$) for $pp$ ($\bar{p}p$)
amplitudes. Using the notation $\alpha_{\tt I\!P}(0) 
= 1+\epsilon$, $\alpha_{+}(0) = 1 -\eta_{+}$ and $\alpha_{-}(0) = 1 -\eta_{-}$ 
for the intercepts of the Pomeron and the $C=+1$ and $C=-1$ 
trajectories, respectively, the total cross sections, Eq. (14), for $pp$ and $\bar{p}p$ 
interactions are 
written as
\begin{eqnarray}
\sigma_{\mathrm{tot}}(s) = X s^{\epsilon} + Y_{+}\, s^{-\eta_{+}} + \tau Y_{-}\, 
s^{-\eta_{-}}.
\end{eqnarray}

\begin{figure}[ht]
\includegraphics[width=7cm,height=7cm]{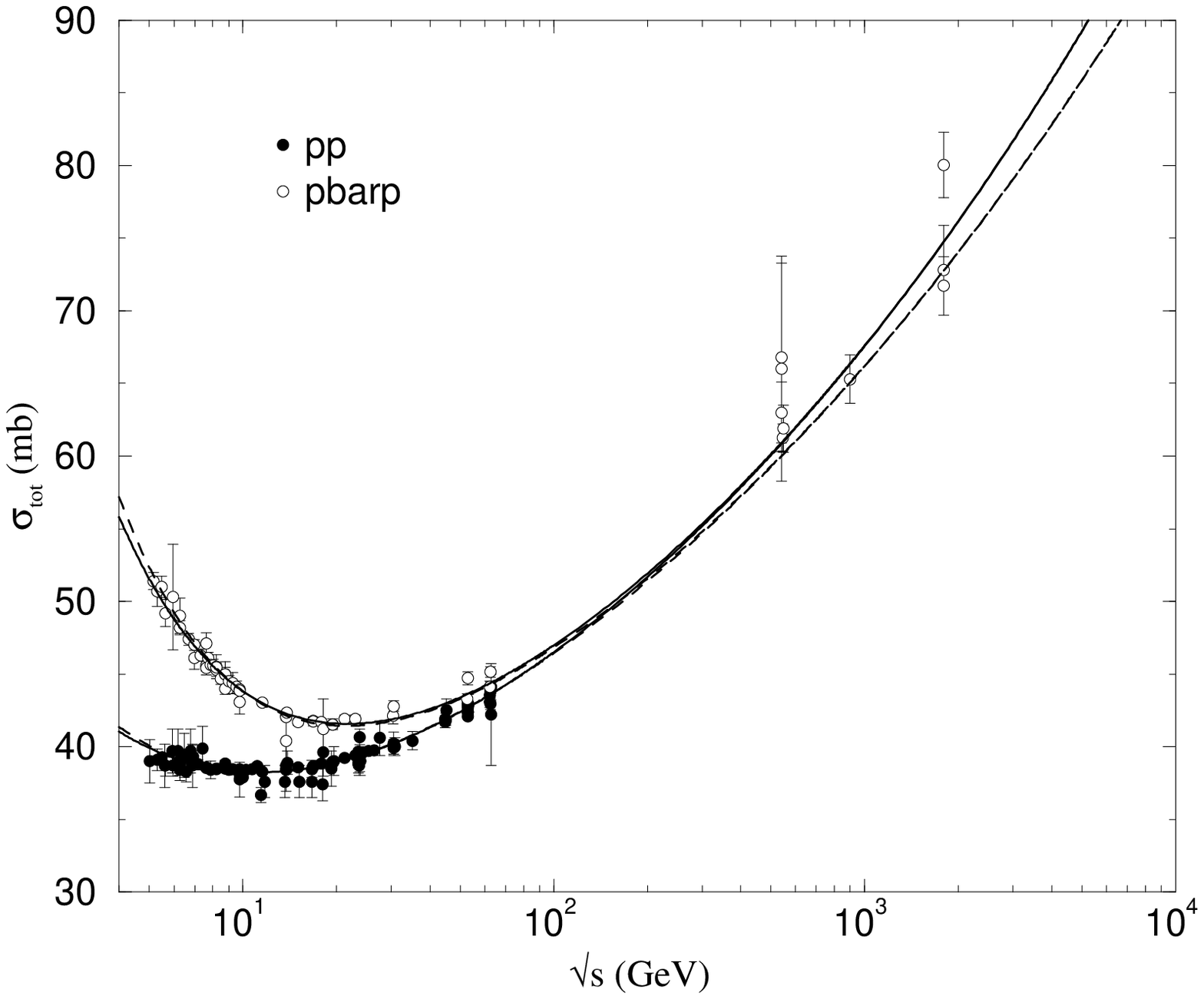}
\includegraphics[width=7cm,height=7cm]{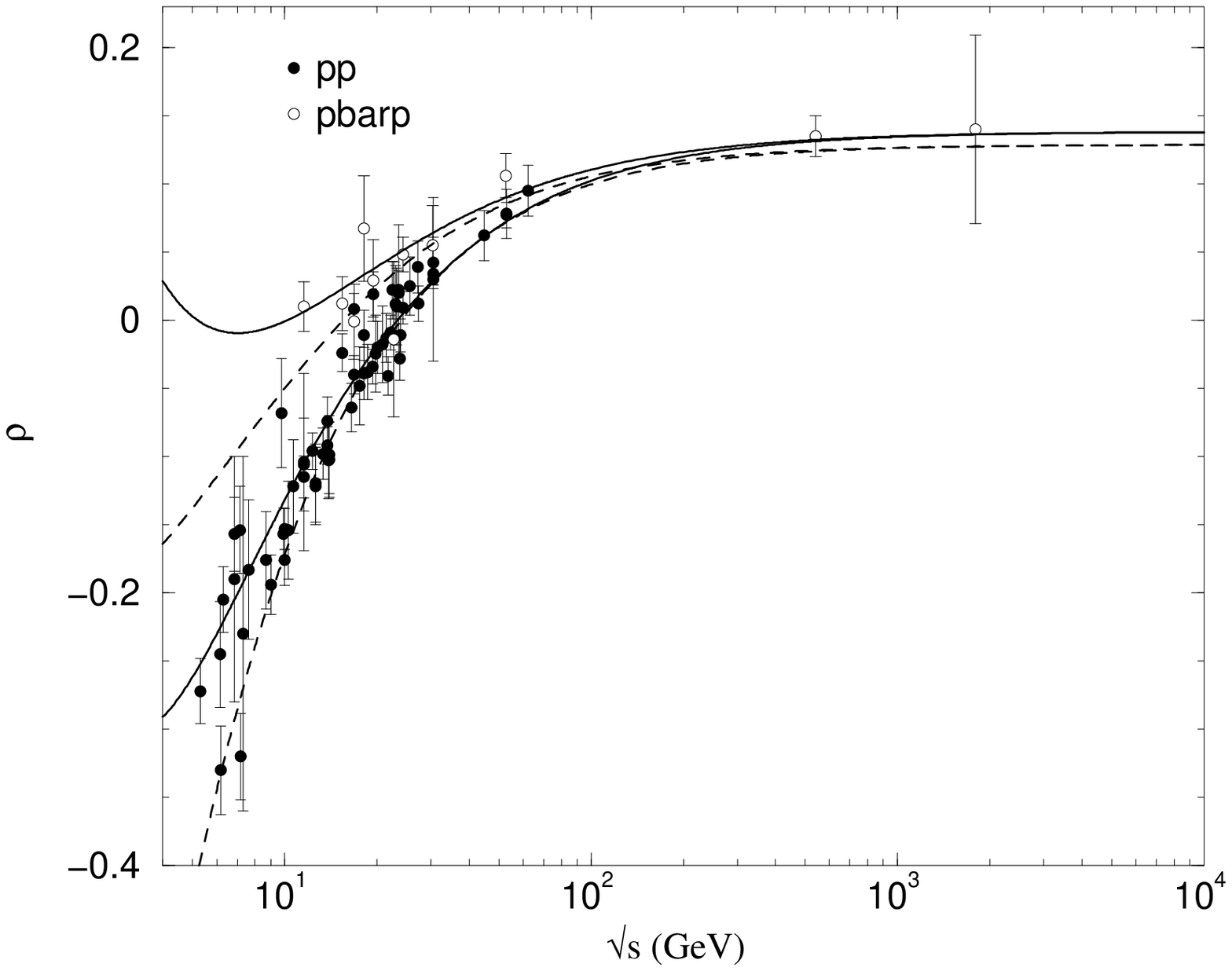}
\caption{Simultaneous fit to $\sigma_{\mathrm{tot}}$ and $\rho$
through the extended parametrization, $\sqrt s_{\mathrm{min}} =$ 5 GeV,
assuming $K=0$ and using either the IDR with $s_0=2m^2$ (solid) or the
DDR (dashed).} \label{fig:5}
\end{figure}

\begin{figure}[ht]
\includegraphics[width=7cm,height=7cm]{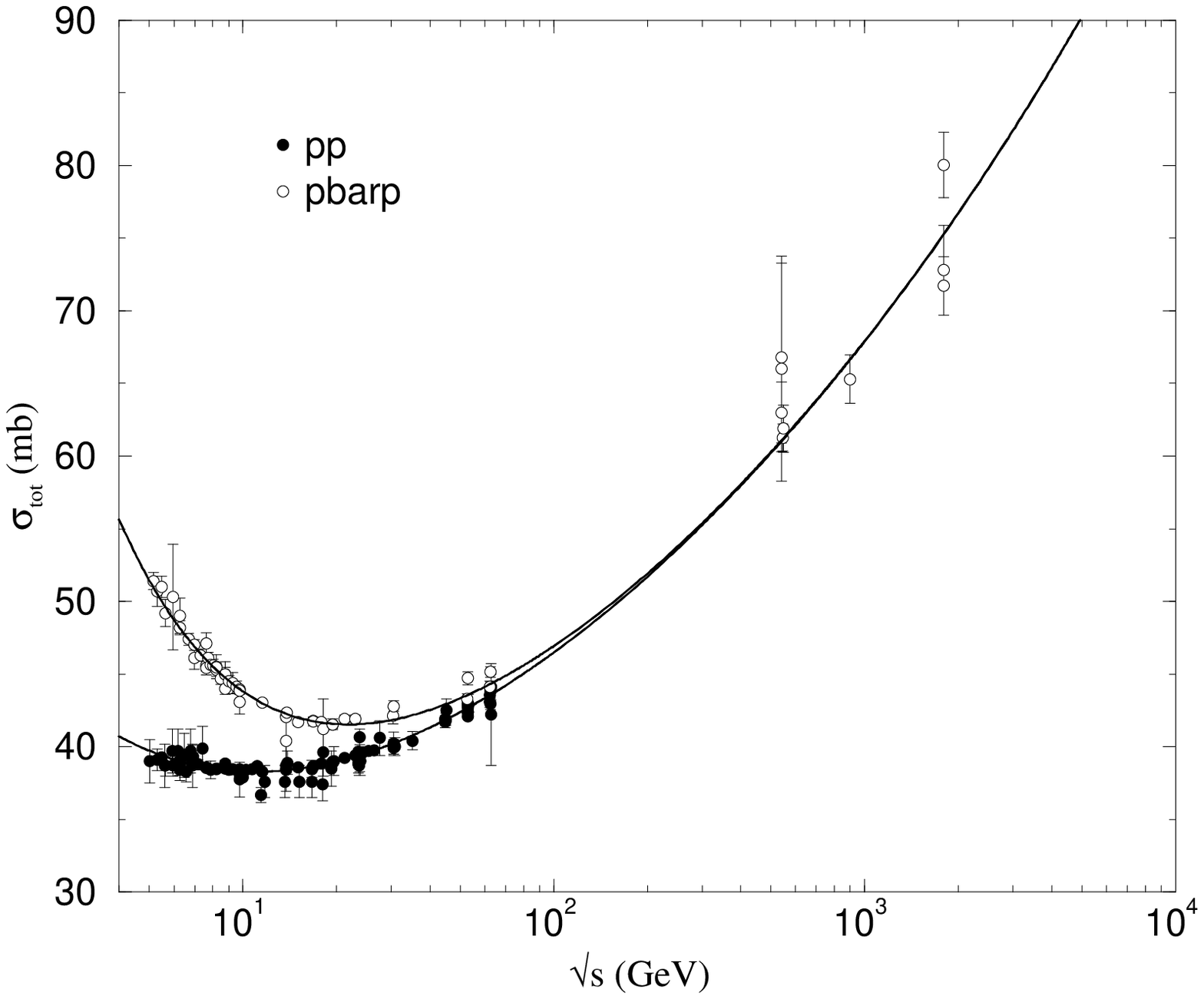}
\includegraphics[width=7cm,height=7cm]{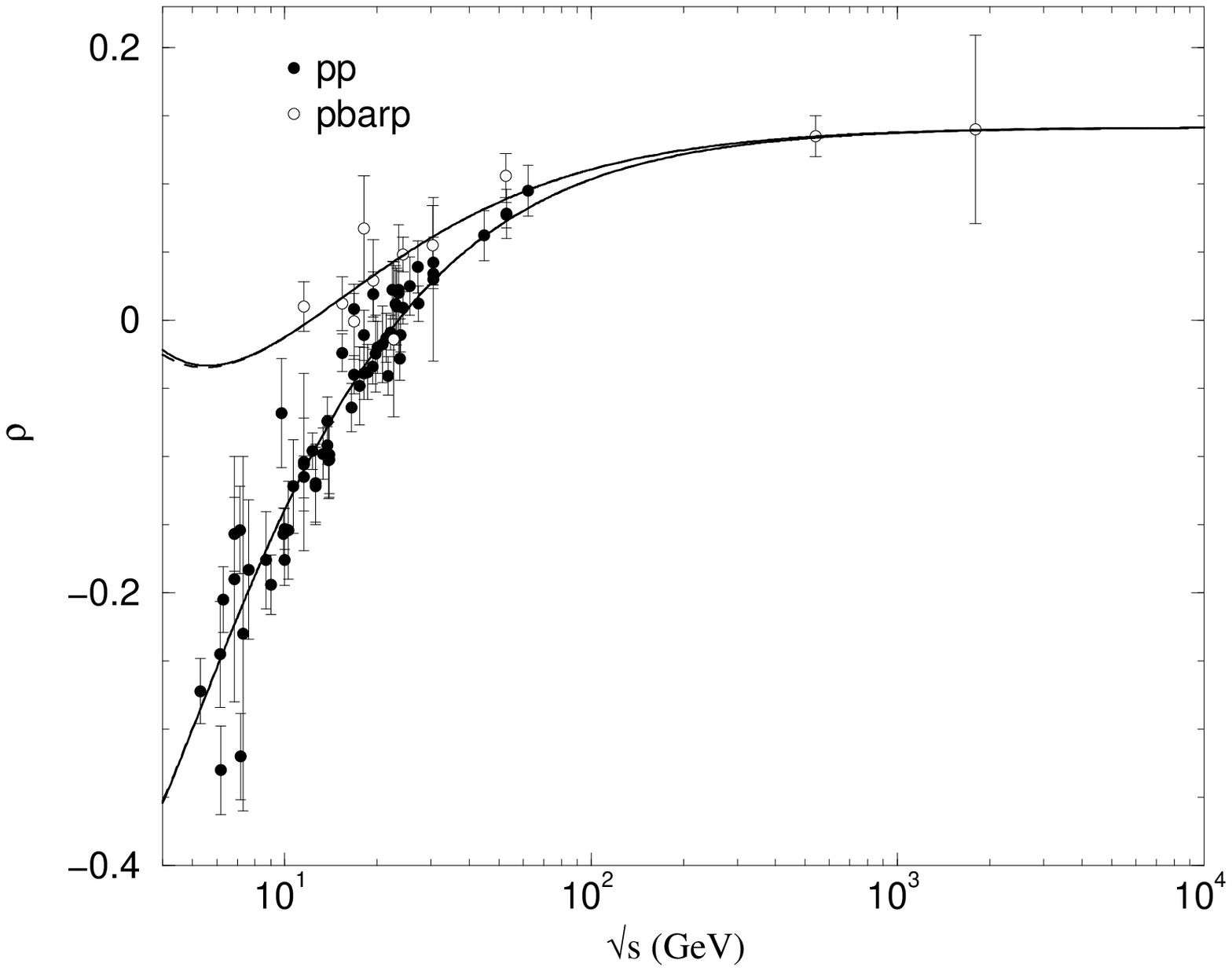}
\caption{Simultaneous fit to $\sigma_{\mathrm{tot}}$ and $\rho$ 
through the extended parametrization, $\sqrt s_{\mathrm{min}} =$ 5 GeV,
with $K$
as free parameter and using either the IDR with $s_0=2m^2$ (solid)
or the DDR (dashed): both curves coincide.} \label{fig:6}
\end{figure}

\begin{figure}[ht]
\includegraphics[width=7cm,height=7cm]{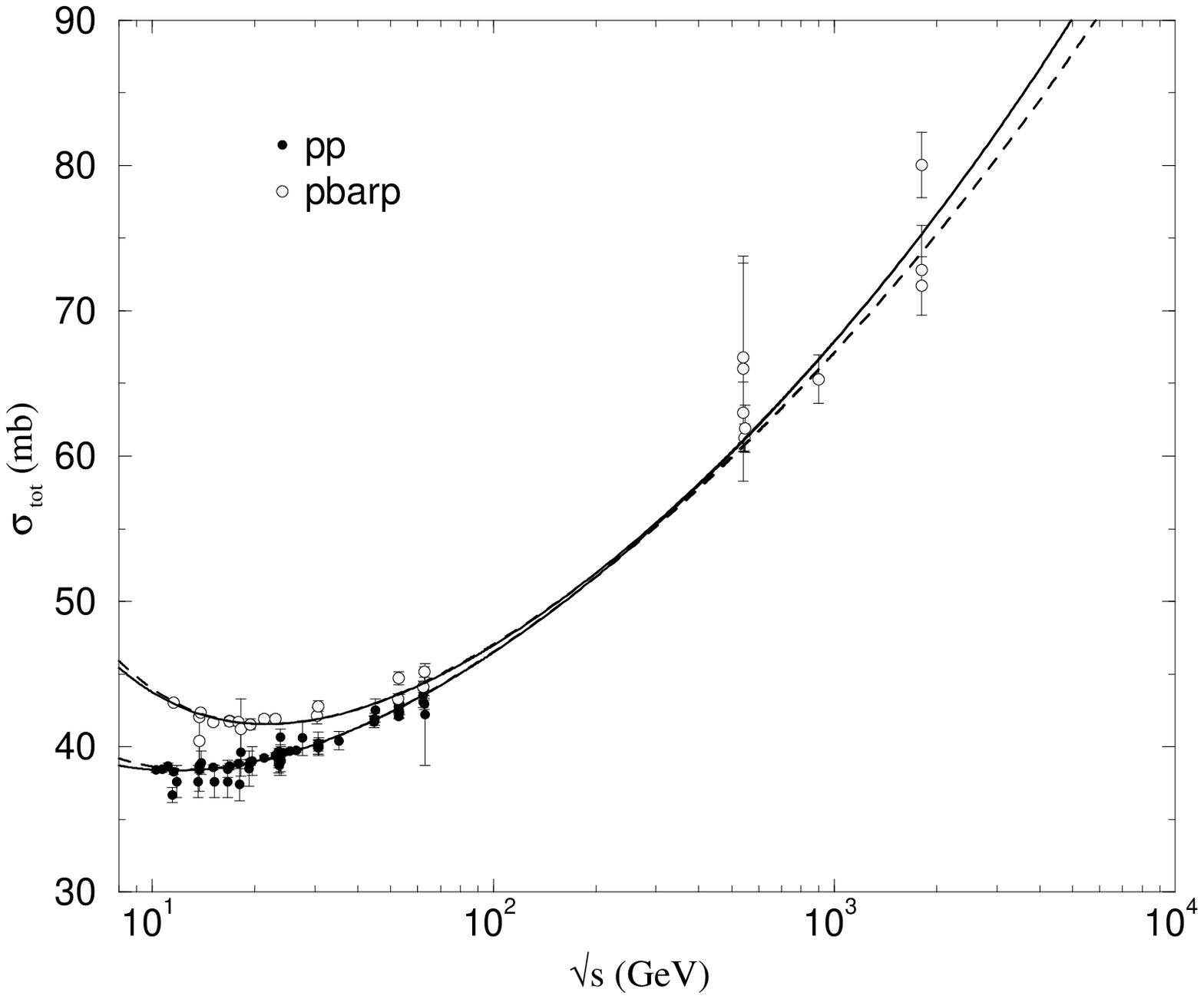}
\includegraphics[width=7cm,height=7cm]{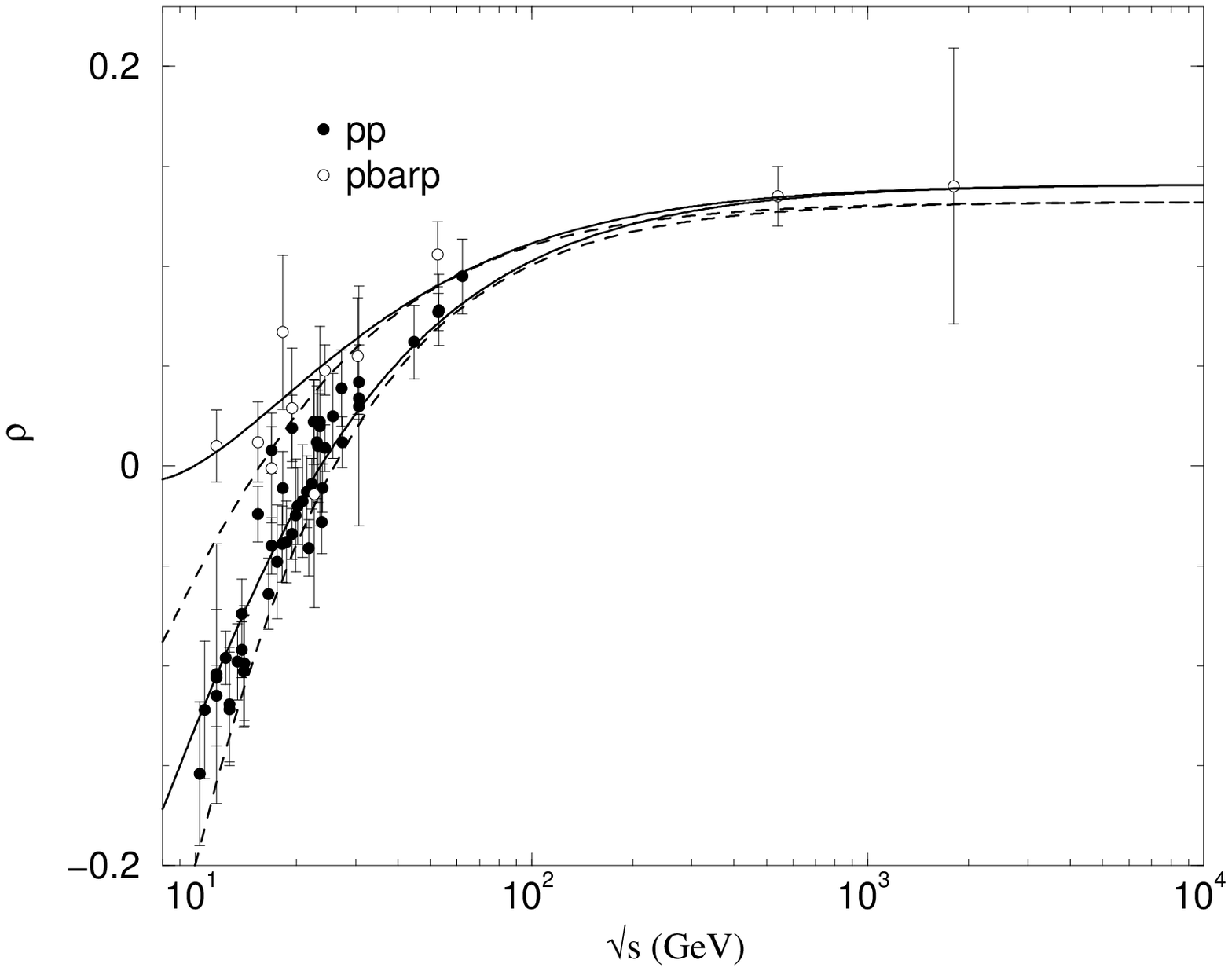}
\caption{Simultaneous fit to $\sigma_{\mathrm{tot}}$ and $\rho$
through the extended parametrization, $\sqrt s_{\mathrm{min}} =$ 10 GeV,
assuming $K=0$ and using either the IDR with $s_0=2m^2$ (solid) or the
DDR (dashed).} \label{fig:7}
\end{figure}

\begin{figure}[ht]
\includegraphics[width=7cm,height=7cm]{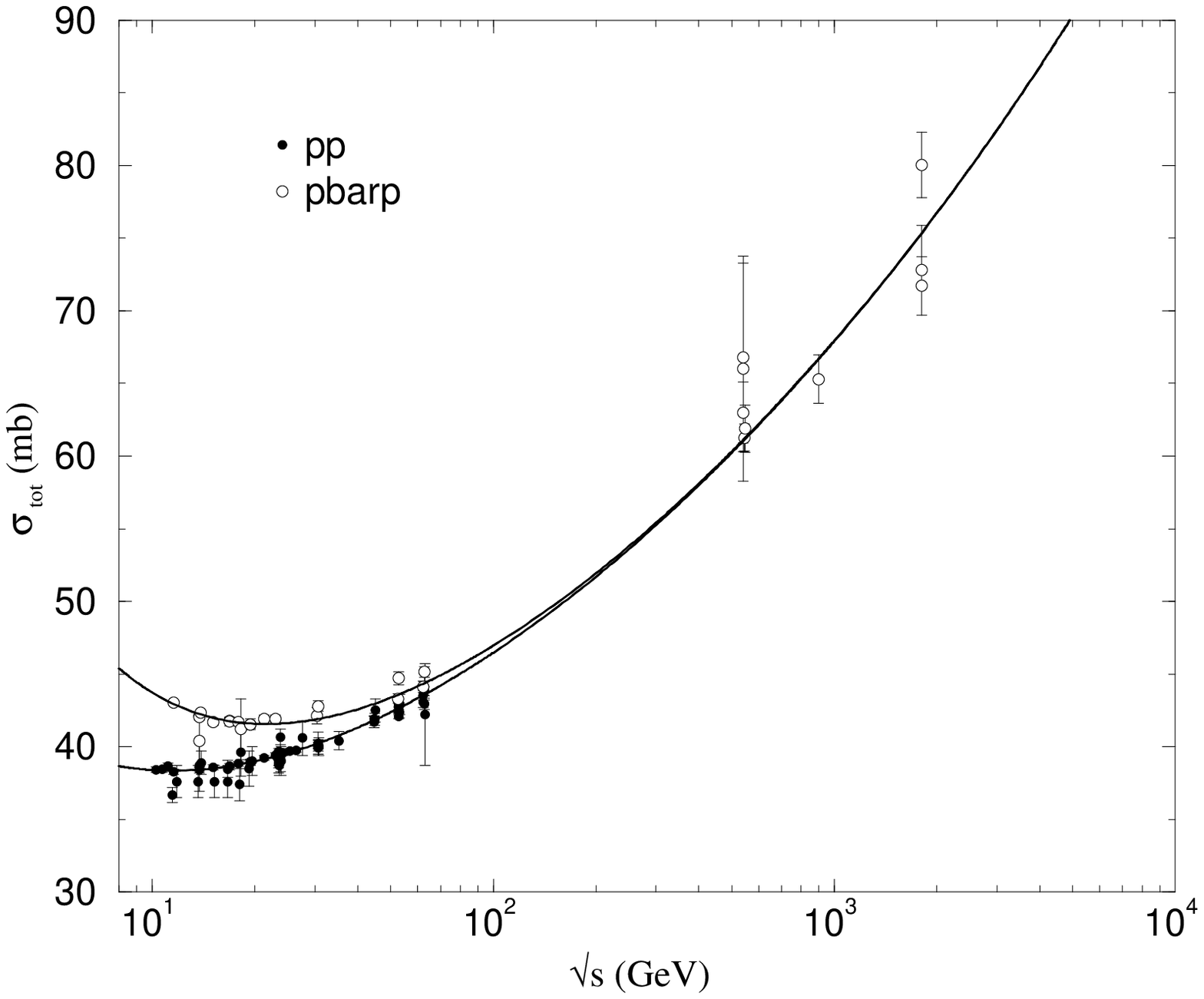}
\includegraphics[width=7cm,height=7cm]{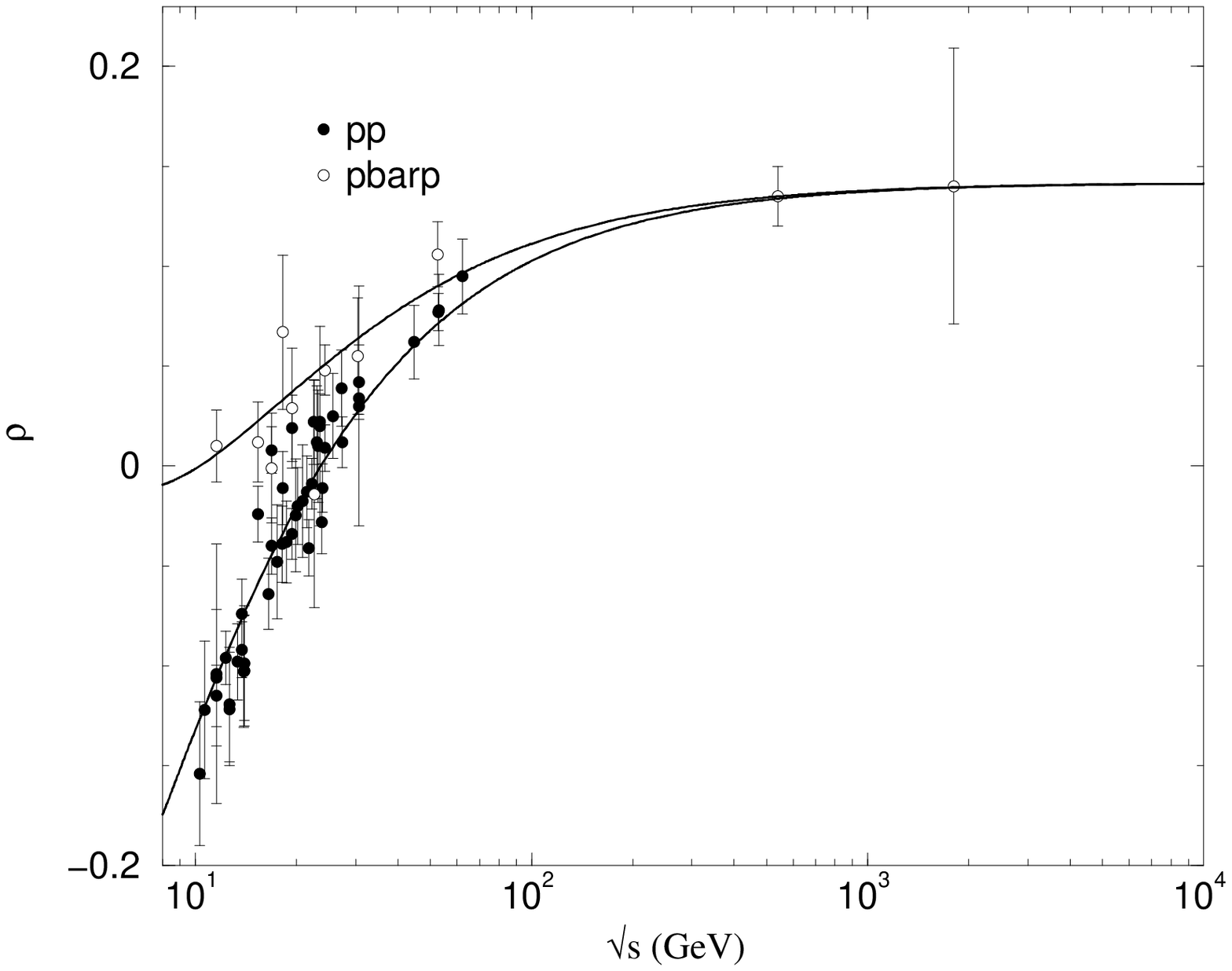}
\caption{Simultaneous fit to $\sigma_{\mathrm{tot}}$ and $\rho$ 
through the extended parametrization, $\sqrt s_{\mathrm{min}} =$ 10 GeV,
with $K$
as free parameter and using either the IDR with $s_0=2m^2$ (solid)
or the DDR (dashed): both curves coincide.} \label{fig:8}
\end{figure}

Since the formal structure of the parametrization is the same as in
the case of degenerate trajectories, with the addition of one more
power term, we shall not display the analytical formulas, but only the
numerical and fit results, together with the experimental data.
The numerical results and statistical information
from all the four cases analyzed with $\sqrt s_{\mathrm{min}} =$ 5 GeV are
 displayed in 
Table \ref{tab:3} and
the curves, together with the experimental data, are shown 
in Figs. \ref{fig:5} and \ref{fig:6}, for $K=0$ or $K$ as a free fit 
parameter, respectively. The corresponding results for 
$\sqrt s_{\mathrm{min}} =$ 10 GeV are presented
in Table \ref{tab:4} and Figs. \ref{fig:7} and \ref{fig:8}.

\begin{table}[h]
\begin{center}
\caption{Simultaneous fits to $\sigma_{\mathrm{tot}}$ and $\rho$
through the extended parametrization,  $\sqrt s_{\mathrm{min}} =$ 5 GeV
(238 data points), either with $K$ as a free
parameter or $K=0$ and using IDR with lower limit $s_0=2m^2$ and DDR.}
\label{tab:3}
\begin{tabular}{ccccc}
\hline
&\multicolumn{2}{c}{IDR with $s_0=2m^2$}&\multicolumn{2}{c}{DDR}\\
\hline
 &    $K$ free         & $K=0$               &     $K$ free          & $K=0$                    \\
\hline

$X$ (mb)   & 19.61  $\pm$  0.53    & 20.12  $\pm$ 0.51    & 19.61  $\pm$ 0.53    & 21.44  $\pm$ 0.44   \\
$Y_+$ (mb) & 66.4   $\pm$  2.0     & 68.9   $\pm$ 2.1     & 66.4   $\pm$ 2.0     & 78.0   $\pm$ 2.0    \\
$Y_-$ (mb) & -33.8  $\pm$  1.9     & -32.4  $\pm$ 1.8     & -33.8  $\pm$ 1.9     & -38.8  $\pm$ 2.6    \\
$\epsilon$ & 0.0895 $\pm$  0.0025  & 0.0874 $\pm$ 0.0024  & 0.0895 $\pm$ 0.0025  & 0.0814 $\pm$ 0.0021 \\
$\eta_+$   & 0.382  $\pm$  0.015   & 0.399  $\pm$ 0.014   & 0.382  $\pm$ 0.015   & 0.450  $\pm$ 0.012  \\
$\eta_-$   & 0.545  $\pm$  0.013   & 0.533  $\pm$ 0.012   & 0.545  $\pm$ 0.013   & 0.573  $\pm$ 0.015  \\
$K$        & -48    $\pm$  15      &   0                  & 69     $\pm$ 18      &   0           \\
$\chi^2/F$ & 1.10       & 1.14                 & 1.10             & 1.64           \\
\hline
\end{tabular}
\end{center}
\end{table}

\vspace{0.5cm}

\begin{table}[h]
\begin{center}
\caption{Simultaneous fits to $\sigma_{\mathrm{tot}}$ and $\rho$ 
through the extended parametrization, $\sqrt s_{\mathrm{min}} =$ 10 GeV
(154 data points), either with $K$ as a free
parameter or $K=0$ and using IDR with lower limit $s_0=2m^2$ and DDR.}
\label{tab:4}
\begin{tabular}{ccccc}
\hline
&\multicolumn{2}{c}{IDR with $s_0=2m^2$}&\multicolumn{2}{c}{DDR}\\
\hline
 &    $K$ free         & $K=0$               &     $K$ free          & $K=0$                    \\
\hline
$X$ (mb)   &   19.57  $\pm$ 0.79   & 19.70 $\pm$ 0.64    & 19.58  $\pm$ 0.78   & 21.088  $\pm$     0.54   \\
$Y_+$ (mb) &   66.0   $\pm$ 6.7    & 67.4  $\pm$ 4.8     & 66.0   $\pm$ 6.6    & 86.850  $\pm$     4.4    \\
$Y_-$ (mb) &   -29.2  $\pm$ 4.0    & -29.1 $\pm$ 3.9     & -29.2  $\pm$ 4.0    & -27.477 $\pm$     6.0    \\
$\epsilon$ &   0.0897 $\pm$ 0.0033 & 0.0892 $\pm$ 0.0028 & 0.0897 $\pm$ 0.0033 & 0.0836  $\pm$     0.0024 \\
$\eta_+$   &   0.380  $\pm$ 0.033  & 0.386 $\pm$ 0.024   & 0.380  $\pm$ 0.033  & 0.46256 $\pm$     0.019  \\
$\eta_-$   &   0.520  $\pm$ 0.025  & 0.519 $\pm$ 0.024   & 0.520  $\pm$ 0.024  & 0.50596 $\pm$     0.040  \\
$K$        &   -14    $\pm$ 48     & 0                   & 104    $\pm$ 58     & 0                        \\
$\chi^2/F$ &   1.10      & 1.09                 & 1.10                 & 1.55\\
\hline
\end{tabular}
\end{center}
\end{table}

\subsection{Discussion}

Making use of power law parametrizations for
the total cross section, we have presented the results of
several simultaneous fits to $\sigma_{\mathrm{tot}}(s)$ and $\rho(s)$
from $pp$ and $\bar{p}p$ scattering.
Once established that the analytical results through IDR with
$s_0$ = 0 and DDR are the same (Section 4.2.1), we have performed
16 fits to the experimental data with the following variants:
(1) degenerate and non-degenerate trajectories;
(2) $\sqrt s_{\mathrm{min}} =$ 5 GeV and $\sqrt s_{\mathrm{min}} =$ 10 GeV;
(3) IDR with $s_0 = 2m^2$ and DDR; 
(4) $K = $ 0 and $K$ as a free
fit parameter. The results are displayed in Tables 1 to 4 and
Figures 1 to 8. In what follows we focus the discussion on the effect of the 
high-energy approximation
($s_0 \rightarrow $ 0 in the IDR leading to the DDR) and on the
role of the subtraction constant.

For $K =$ 0, the only difference between the IDR and the DDR approaches is 
associated with the high-energy approximation, analytically represented by 
integrations of Eqs. (4) and (5) from $s' = 0$ to $s' = s_0 = 2m^2 \approx $
 1.8 GeV.
In principle, it might be expected that this contribution could affect the
fit results only in the low-energy region, roughly below $\sqrt s \approx$
50 GeV. However, the fit procedure is characterized by strong correlations 
among the free parameters and, therefore, the contributions from any
region may be ``communicated"to other regions and that is exactly what
our results show. In fact, for $K =$ 0, independently
of the parametrization and the lower energy cut used, the leading 
contribution at high energies (represented by the value of the Pomeron
intercept, $\epsilon$) is different when using IDR or DDR. The above 
approximation (DDR) results in a value for the intercept that is smaller 
than that obtained with the IDR. Specifically, in the case of the DL
parametrization (Tables 1 and 2), the reduction reads 0.67 \% for
$\sqrt s_{\mathrm{min}} =$ 5 GeV and 0.7 \% for $\sqrt s_{\mathrm{min}} =$ 10 GeV
(see $\sigma_{\mathrm{tot}}(s)$ in Figs 1 and 3). The effect is more significant
in the case of the extended parametrization (Tables 3 and 4),
with a reduction of 6.9 \% for $\sqrt s_{\mathrm{min}} =$ 5 GeV and 6.3 \% for 
$\sqrt s_{\mathrm{min}} =$ 10 GeV
(see $\sigma_{\mathrm{tot}}(s)$ in Figs. 5 and 7).
That is an important and novel result, which shows that, in practice, for the
simple-pole Pomeron and secondary reggeons, the high-energy approximation
enclosed in the DDR affects the fits results even at the asymptotic energies.
As mentioned, that is a consequence of the fit procedure 
and the correlations among the
free parameters.

Now, the practical role of the subtraction constant may be investigated
by comparing the results obtained with $K$ = 0 and those with $K$ as
a free fit parameter. The striking result here is the fact that,
independently of the parametrization and the energy cut used, the numerical
results obtained for the fit parameters with the IDR and with the
DDR are the same up to 3 significant figures, including the $\chi^2/F$. 
For example, from Tables 1 and 2, for the case of the DL parametrization, we
have obtained in both cases (IDR with $s_0 = 2m^2$ and DDR) the same values: 
$\epsilon = 0.0747 \pm 0.0014$, $\chi^2/F$ = 1.39 
for $\sqrt s_{\mathrm{min}} =$ 5 GeV
and
$\epsilon = 0.0816 \pm 0.0018$, $\chi^2/F$ = 1.17 
for $\sqrt s_{\mathrm{min}} =$ 10 GeV
(see Figs 2 and 4, respectively).
For the cases of the extended parametrization, Tables 3 and 4, we have 
obtained
$\epsilon = 0.0895 \pm 0.0025$, $\chi^2/F$ = 1.10 
for $\sqrt s_{\mathrm{min}} =$ 5 GeV
and
$\epsilon = 0.0897 \pm 0.0033$, $\chi^2/F$ = 1.10 
for $\sqrt s_{\mathrm{min}} =$ 10 GeV
(see Figs 6 and 8, respectively).

All these numerical results can be understood in an analytical
context if we investigate the correlation between the subtraction constant
and the contribution below the lower integration limit 
(referred to in Sec. 4.2.1). To this end, let us return to the product
of $\rho(s)$ by $\sigma_{tot}(s)$ in the case of IDR with $s_0 = 2m^2$,
as given by Eq. (18), or the corresponding formula with non-degenerate 
trajectories. 
In both cases the terms with entire inverse powers of $s$ can be expressed by

\begin{eqnarray}
[ K + \Delta ] \frac{1}{s} +
{\mathcal{O}}( 1/s^2), \nonumber
\end{eqnarray}
where $K$ is the subtraction constant and $\Delta$ comes from the
series expansion, which is associated with the integration from $s' = 0$ to
$s' = 2m^2$ (Sec. 4.2.1). Therefore, at large enough energies, this
contribution below the lower integration limit is expected to be
absorbed in the subtraction constant. In fact, with degenerate
(de) and non-degenerate (non-de) trajectories, these contributions read

\begin{eqnarray}
\Delta_{\mathrm{de}} =
\frac{1}{\pi}
\left(
\frac{2Xs{_0}^{1+\epsilon}}{1+\epsilon}
+\frac{(Y+Z)s{_0}^{1-\eta}}{1-\eta}
\right), \nonumber
\end{eqnarray}

\begin{eqnarray}
\Delta_{\mathrm{non-de}} =
\frac{1}{\pi}
\left(
\frac{2Xs{_0}^{1+\epsilon}}{1+\epsilon}
+\frac{2Y_{+}s{_0}^{1-\eta_{+}}}{1-\eta_{+}}
\right), \nonumber
\end{eqnarray}
and may be estimated with the central values of the fit parameters
from Tables 1-4:

$\Delta_{\mathrm{de}} \approx$ 162 \ ($\sqrt s_{\mathrm{min}} = $ 5 GeV),
$\qquad$
$\Delta_{\mathrm{de}} \approx$ 170 \ ($\sqrt s_{\mathrm{min}} = $ 10 GeV),

$\Delta_{\mathrm{non-de}} \approx$ 118 \ ($\sqrt s_{\mathrm{min}} = $ 5 GeV),
$\qquad$
$\Delta_{\mathrm{non-de}} \approx$ 117 \ ($\sqrt s_{\mathrm{min}} = $ 10 GeV),

From those Tables, we see that these values are in complete agreement with 
the differences
between the fitted values of the subtraction constants determined with DDR
and IDR for $s_0 = 2m^2$. Therefore, we conclude that in the case of
DDR we have an ``effective''
subtraction constant which can compensate (analytically and numerically)
the effect of the high-energy approximation.

That is another important and novel result, showing explictly the
practical role of this parameter: for simple-pole pomeron and secondary reggeons
once the subtraction constant is considered as
a free fit parameter, the DDR are completely equivalent
to the IDR with fixed (non-zero) lower limit.
We also note that our qualitative and quantitative conclusions about
the effects of the high-energy approximation and the subtraction constant
are the same with both the degenerate and non-degenerate trajectories
and both energy cuts, 5 and 10 GeV.

To conclude this Section, let us call attention to some critical aspects 
related with the use of the DL and the extended parametrizations. 
From a statistical point of view, the extended parametrization leads to a better
description of the experimental data (Tables 1-4).
Also, the
$\rho$ data at 546 GeV and 1.8 TeV are well described with the
extended parametrization, but not in the
DL case.
The essential difference between these two parametrizations concerns
the additional secondary reggeon, represented by the dependence
$s^{-\gamma}$, 0 $< \gamma <$ 1, which goes to zero as the energy increases. 
We conclude that the splitting of the trajectories allows the free parameters from
the secondary reggeons to fit the data at low energies (between 5 GeV and,
let us say, 100 GeV), giving more freedom for the parameters associated
with the Pomeron to fit the data at the highest energies. That seems to be
well known and also well accepted in the literature \cite{nondege}.
However, we can not forget that the Regge phenomenology is intended only
for asymptotic energies. Specifically, the basic contribution 
comes from the asymptotic form of the Legendre Polynomial, namely 
$P_{l}(x) \rightarrow x^{l} \quad \textrm{as} \quad
x \rightarrow \infty$,
where $x$ is translated to the energy $s$ through the crossing symmetry
and $l$ to the trajectory of the Pomeron or Reggeon \cite{ddln}.
On the other hand, from the above discussion, the secondary reggeons
act in a region where the total cross section reaches its
minimum (followed by an increase) and the $\rho$ parameter cross the
zero (becoming positive). Even if we, optimistically, concentrate in
the above region, $\sqrt s =$ 5 to 100 GeV, that region, certainly, has nothing 
to do with an asymptotic concept.
Based on these facts, we understand that, although the secondary reggeons
can be used as suitable parametrizations on statistical or strictly fit
grounds, one must be careful in attempting to
extract well founded physical results from these
asymptotic forms used at so low values of the energy. 
On the other hand,
that is not the case for the subtraction constant as a free fit parameter,
since it has well defined mathematical bases, which are related with
polynomial bounds.

\section{Conclusions and final remarks} \label{sec:conclusions}

In this work we have presented a review on the different results and statements
in the literature related with the replacement of IDR by DDR, and a discussion 
connecting these different aspects with the corresponding assumptions and
classes of functions considered in each case.

By means of a formal and analytical approach, we have demonstrated that
the subtraction constant is preserved when the IDR are replaced by the
DDR and that, for the class of functions entire in $\ln s$, the DDR do
not depend on any additional free parameter (except for the subtraction
constant). We have stressed that the only approximation involved in this
replacement concerns the lower limit $s_0$ in the IDR: the high-energy
condition is reached by assuming that $s_0 = 2m^2 \rightarrow 0$.

We have investigated the practical applicability of the DDR and IDR in the
context of the Pomeron-reggeon parametrizations, with both degenerate and
non-degenerate higher meson trajectories. By means of global fits to
$\sigma_{\mathrm{tot}}(s)$ and $\rho(s)$ data from $pp$ and $\bar{p}p$ scattering,
we have tested all the important variants that could affect the fit
results, namely the number of secondary reggeons, energy cutoff, effects of
the high-energy approximation and the subtraction constant and the derivative
approach with DDR and IDR with fixed $s_0$. Our results lead to the conclusion that
the high-energy approximation and the subtraction constant affect the
fit results at both low and high energies. This effect is a consequence of the 
fit procedure, associated with the strong correlation among the free parameters.

A striking novel result concerns the practical role of the subtraction
constant. We have shown that, with the Pomeron-reggeon parametrization,
once the subtraction constant is used as a free fit parameter, the results 
obtained with the DDR and with the IDR (with finite lower limit, $s_0 = 2m^2$)
are the same up to 3 significant figures in the fit parameters and
$\chi^2/F$. Analytically this effect is due to the absorption of the
high-energy approximation in an ``effective'' subtraction constant.
This conclusion, as we have shown, is independent of the
number of secondary reggeons (DL or extended parametrization) or the
energy cutoff ($\sqrt s$ = 5 or 10 GeV).

We have called the attention to the fact that the subtraction constant has well 
founded mathematical bases, since it is a consequence of the polynomial bounds
in the scattering amplitude. On the other hand, the use of the asymptotic
Regge forms for detailed fits at finite energies 
(minimum on $\sigma_{\mathrm{tot}}(s)$) is not formally justified.
We understand that our results suggest that, presently, it is very important 
to look for
usefull and well founded theoretical results at finite energies
and that one must be careful in attempting
to predict asymptotic behaviors based on asymptotic
formalisms applied at finite energies.

In order to treat a complete example, with the 16 variants referred to in the text,
we have considered here only a simple-pole representation for the Pomeron.
Although belonging to the class of functions entire in $\ln s$, this choice is
outside the more general class of functions considered by 
Fischer and P. Kol\'a\v{r}, because it violates the Froissart-Martin bound at
the asymptotic energies. We are presently investigating other possibilities,
as dipole and tripole contributions. 
Anyway, there is an interesting aspect 
concerning these forms, which we shall mention here. The possibility that
the odd amplitude had a $\ln s$ contribution (Odderon) has been
investigated in Refs. \cite{kn,alm03}, by means of the DDR. However, this 
contribution in the IDR, Eq. (5), with $s_0 $ = 0 is divergent. On the
other hand, that is not the case if we use the DDR as expressed by Eqs. (9)
or (13). Since we have shown that the only approximation in going from
IDR to DDR is to take $s_0 \rightarrow 0$ in the integral relation,
the above results seems inconsistent. We think that this may be associated with
the formal substitution of the tangent by the cotangent operator, as referred
to in Sec. $3.1$. We are presently investigating this subject.

The DDR introduced by Bronzan, Kane, and Sukhatme \cite{bks},
 and later generalized for an
arbitrary number of subtractions by Menon, Motter and Pimentel
\cite{mmp} bring enclosed the parameter $\alpha$, which we have shown
does not need to take part in the calculation. We understand that this 
dependence is not
necessarily wrong if, in practical uses, one takes $\alpha = 1$.
Therefore, the presence of this parameter is only an unnecessary
complication. Certainly, considering $\alpha$ as a free parameter
at finite energies may improve the fit results \cite{bks,mmpsn},
but it seems to us difficult to justify its mathematical and,
mainly, its physical meaning.

{\em Note added in proof}. After the submission of this work, the effect
of the absorption of the high-energy approximation ($s_0 \rightarrow$ 0)
by the subtraction constant has been confirmed by other authors
\cite{cudell03}.

\begin{ack}
We are thankful to Prof. M. Giffon and A.F. Martini for useful
information on
the use of hypergeometric functions, and also to A. Maia Jr.,
E.C. Oliveira, E.G.S. Luna and J. Montanha for fruitfull discussions.
We are grateful to two anonymous referees for valuable criticisms and
comments.
This work has been supported by FAPESP
(Contract N. 03/00228-0 and 00/04422-7).
\end{ack}


\begin{thebibliography}{99}

\bibitem{drgeneral}
R.J. Eden,
High Energy Collisions of Elementary Particles,
Cambridge University Press, Cambridge, 1967;
H.M. Nussenzveig,
Causality and Dispersion Relations,
Academic Press, New York, 1972;
P.D.B. Collins,
An Introduction to Regge Theory and High Energy Physics,
Cambridge University Press, Cambridge, 1977.

\bibitem{bc}
M.M. Block, R.N. Cahn, Rev. Mod. Phys. 57 (1985) 563.

\bibitem{idr}
M.L. Goldberger, Y. Nambu, R. Oehme, Ann. Phys. 2 (1957) 226;
P. S\"oding, Phys. Lett. 8 (1964) 285.


\bibitem{bourrely73}
C. Bourrely, J. Fischer, Nucl. Phys. B 61 (1973) 513.

\bibitem{amaldi77}
U. Amaldi et al., Phys. Lett. B 66 (1977) 390.

\bibitem{kluit}
P.M. Kluit, J. Timmermans, Phys. Lett. B 202 (1988) 458.

\bibitem{ua42}
UA4/2 Collaboration, C. Augier et al., Phys. Lett. B 315 (1993) 503.

\bibitem{kvw}
K. Kang, P. Valin, A.R. White, Il Nuovo Cimento A 107 (1994) 2103.

\bibitem{bertini}
M. Bertini, M. Giffon, L. Jenkovszky, F. Paccanoni, Il Nuovo Cimento A
109 (1996) 257.

\bibitem{gm}
V.N. Gribov, A.A. Migdal, Yad. Fiz. 8  (1968) 1002 [Sov. J. Nucl.
Phys. 8 (1969) 583]; Yad. Fiz. 8 (1968) 1213
[Sov. J. Nucl. Phys. 8 (1969) 703].


\bibitem{bronzan68}
J.B. Bronzan, in: Symposium on the Pomeron,
Argonne National Laboratory,
ANL/HEP-7327 (1973) 33.

\bibitem{jack73}
 J.D. Jackson, in: R.L. Crawford, R. Jennings (Eds.),
Proc. Fourteenth Scottish Universities Summer School in Physics,
Phenomenology of Particles at High Energies,
Academic Press, London, 1974, pp. 1--103.

\bibitem{bks}
J.B. Brozan, G.L. Kane, U.P. Sukhatme, Phys.
Lett. B 49 (1974) 272.


\bibitem{kn}
K. Kang, B. Nicolescu, Phys. Rev. D 11
(1975) 2461.

\bibitem{cudelletall}
J.R. Cudell, K. Kang, S.K. Kim, Phys. Lett. B 395 (1997) 311;
J.R. Cudell, V.V. Ezhela, K. Kang, S.B. Lugousky, N.P. Tkachenko, Phys. Rev. D
61 (2000) 034019; 63 (2001) 059901 (erratum).


\bibitem{competework}
COMPETE Collaboration, J.R. Cudell et al., Phys. Rev. D 65
(2002) 074024; Phys. Rev. Lett. 89 (2002) 201801.


\bibitem{crit1}
G. K. Eichmann, J. Dronkers, Phys. Lett. B 52 (1974) 428.

\bibitem{crit2}
J. Heidrich, E. Kazes, Lettere al Nuovo Cimento 12
(1975) 365.

\bibitem{crit3}
G. H\"ohler, H.P. Jakob, F. Kaiser, Phys. Lett. B 58 (1975) 348.

\bibitem{crit4}
 A. Bujak, O. Dumbrajs, J. Phys. G: Nucl. Phys. 12
(1976) L129.


\bibitem{kf}
J. Fischer, P. Kol\'a\v{r}, Phys. Lett. B 64
(1976) 45; Phys. Rev. D 17 (1978) 2168; P. Kol\'a\v{r}, J. Fischer, J.
 Math. Phys. 25 (1984) 2538;
J. Fischer, P. Kol\'a\v{r}, Czech. J. Phys. B 37
(1987) 297.

\bibitem{vrkoc}
I. Vrko\v{c}, Czech. Math. J. 35 (1985) 59.

\bibitem{mmp}
M.J. Menon, A.E. Motter, B.M. Pimentel, Phys. Lett. B 451  (1999) 207.

\bibitem{mmpsn}
A.F. Martini, M.J. Menon, J.T.S. Paes, M.J. Silva Neto, Phys. Rev. D
59 (1999) 116006; R.F. \'Avila, E.G.S. Luna, M.J. Menon, Braz. J.
Phys. 31 (2001) 567.

\bibitem{kfrev}
P. Kol\'a\v{r}, J. Fischer, in: V. Kundrat and P. Zavada (Eds.), Proc. Blois
Workshop on Elastic and
Diffractive Scattering, IOP, Prague, 2002,
pp. 305--312.

\bibitem{alm03}
R.F. \'Avila, E.G.S. Luna, M.J. Menon, Phys. Rev. D 67
(2003) 054020; in C.A.Z. Vasconcelos et al. (Eds.), Structure and
Interactions of
Hadronic Systems, World Scientific, Singapore, 2003,
p. 421--424.

\bibitem{lm03}
E.G.S. Luna, M.J. Menon, Phys. Lett. B 565 (2003) 123.


\bibitem{am02}
R.F. \'Avila, M.J. Menon, in: Y. Hama and F.S. Navarra (Eds.),
Relat\'orio da 14a. Reuni\~ao de Trabalho em
Intera\c c\~oes Hadr\^onicas,
IFUSP, S\~ao Paulo, 2002, p. 90 - 95; in: XXIII Brazilian National
Meeting on Particles and Fields, 2002, SLAC SPIRES-HEP, Aguas de Lindoia
Server,
www.sbf1.if.usp.br/eventos/enfpc/xxiii/procs/RES28.


\bibitem{cms}
J.R. Cudell, E. Martynov, O. Selyugin, hep-ph/0307254.

\bibitem{odd}
L. Lukaszuk. B. Nicolescu, Nuovo Cimento Lett. 8, 405 (1973).

\bibitem{pc}
B. Nicolescu (private communication).

\bibitem{suketal}
U. Sukhatme, G.L. Kane, R. Blankenbecler, M. Davier,
Phys. Rev. D 12, 3431 (1975).

\bibitem{sidhu}
D.P. Sidhu, U.P. Sukhatme, Phys. Rev. D 11, 1351 (1975).

\bibitem{ft}
A. Martin, Scattering Theory: Unitarity, Analyticity
and Crossing, Springer, New York, 1969; R.J. Eden,
Rev. Mod. Phys. 43, 15 (1971); J. Fischer, Phys. Rep. 76, 157 (1981).

\bibitem{ddln}
S. Donnachie, G. Dosch, P. Landshoff, O. Nachtmann, Pomeron
Physics in QCD, Cambridge University Press, Cambridge, 2002.

\bibitem{pdg}
Particle Data Group, K. Hagiwara et al. Phys. Rev. D 66 (2002) 010001,
the full data sets are available at
http://pdg.lbl.gov.



\bibitem{dl}
A. Donnachie, P. V. Landshoff, Z. Phys. C 2 (1979) 55;
Phys. Lett. B 387 (1996) 637;
Phys. Lett. B 296  (1992) 227.


\bibitem{tables}
I.S. Gradshteyn, I. M. Ryzhik,
Table of Integrals, Series and Products,
Academic Press, San Diego, 1980;
M. Abramowitz, I. A. Stegun,
Handbook of Mathematical Function,
Dover, New York, 1964.

\bibitem{nondege}
R.J.M. Covolan, J. Montanha , and K. Goulianos, Phys. Lett. B 389 (1996) 176;
J.R. Cudell, K. Kang, and S.K. Kim, Ref. \cite{cudelletall}.

\bibitem{cudell03}
J.R. Cudell, E. Martynov, O. Selyugin, A. Lengyel, Phys. Lett. B 587
(2004) 78; E. Martynov, J.R. Cudell, O. Selyugin,
hep-ph/0311019.

\end{thebibliography}
\end{document}